\documentclass[prd,preprint,tightenlines,preprintnumbers,aps,eqsecnum,nofootinbib,superscriptaddress,floatfix]{revtex4-1}

\usepackage{amsmath}    
\usepackage{amssymb}
\usepackage{graphicx}   
\usepackage{verbatim}   
\usepackage{color}      
\usepackage{subfigure}  
\usepackage{longtable}
\usepackage{siunitx}
\usepackage[colorlinks=true,citecolor=blue,linkcolor=blue,urlcolor=blue]{hyperref}

\usepackage{bm}         
\let\mathbf=\bm         

\usepackage{dsfont}

\newcommand\T{\rule{0pt}{2.6ex}}
\newcommand\B{\rule[-1.2ex]{0pt}{0pt}}

\newcommand{\be}{\begin{equation}}
\newcommand{\ee}{\end{equation}}
\newcommand{\bea}{\begin{eqnarray}}
\newcommand{\eea}{\end{eqnarray}}

\newcommand{\crit}{\ensuremath{\kappa_\text{crit}}}
\newcommand{\order}{\ensuremath{\mathrm{O}}}

\renewcommand{\case}[2]{\ensuremath{{\textstyle\frac{#1}{#2}}}}

\newcommand{\quarter}{\ensuremath{\case{1}{4}}}
\newcommand{\oneninth}{\ensuremath{\case{1}{9}}}

\definecolor{green}{rgb}{0.0,0.6,0.0}
\definecolor{blue}{rgb}{0.0,0.0,0.6}
\definecolor{gray}{rgb}{0.6,0.6,0.6}

\newcommand{\cut}[1]{}

\begin{document}

\title{Splittings of low-lying charmonium masses at the physical point}

\author{Carleton DeTar}
\email{detar@physics.utah.edu}
\affiliation{Department of Physics and Astronomy, University of Utah, \\ Salt Lake City, Utah, USA}

\author{Andreas~S.~Kronfeld}
\email{ask@fnal.gov}
\affiliation{Fermi National Accelerator Laboratory, Batavia, Illinois 60510-5011, USA}
\affiliation{Institute for Advanced Study, Technische Universit\"at M\"unchen, Garching, Germany}

\author{Song-haeng~Lee}
\affiliation{Department of Physics and Astronomy, University of Utah, \\ Salt Lake City, Utah, USA}

\author{Daniel Mohler}
\email{damohler@uni-mainz.de}
\affiliation{Helmholtz-Institut Mainz, 55099 Mainz, Germany}
\affiliation{Johannes Gutenberg-Universit\"at Mainz, 55099 Mainz, Germany}

\author{James~N.~Simone}
\email{simone@fnal.gov}
\affiliation{Fermi National Accelerator Laboratory, Batavia, Illinois 60510-5011, USA}

\collaboration{Fermilab Lattice and MILC Collaborations}
\preprint{FERMILAB-PUB-18-440-T}
\preprint{MITP/18-097}
\noaffiliation

\date{\today}

\begin{abstract}
We present high-precision results from lattice QCD for the mass splittings of the low-lying charmonium states.
For the valence charm quark, the calculation uses Wilson-clover quarks in the Fermilab interpretation.
The gauge-field ensembles are generated in the presence of up, down, and strange sea quarks, based on the improved staggered
(asqtad) action, and gluon fields, based on the one-loop, tadpole-improved gauge action.
We use five lattice spacings and two values of the light sea quark mass to extrapolate the results to the physical point.
An enlarged set of interpolating operators is used for a variational analysis to improve the determination of the energies of the
ground states in each channel.
We present and implement a continuum extrapolation within the Fermilab interpretation, based on power-counting
arguments, and thoroughly discuss all sources of systematic uncertainty.
We compare our results for various mass splittings with their experimental values, namely, the 1S hyperfine splitting, the 1P-1S
splitting and the P-wave spin-orbit and tensor splittings.
Given the uncertainty related to the width of the resonances, we find excellent agreement.
\end{abstract}


\maketitle

\section{Introduction}
\label{sec:intro}

Over the past decade, the experimental study of the products of $B$-meson decays has led to the discovery of a wealth of excited
charmonium states.
Many of them present interesting challenges for theoretical interpretation.
Because lattice quantum chromodynamics (QCD) is an {\it ab initio} method for studying hadron spectroscopy, in principle, it should
provide a guide to the interpretation of these states~\cite{Dudek:2007wv,Burch:2009az,Bali:2009er,Bali:2011rd,Bali:2012ua,%
Liu:2011rn,Namekawa:2011wt,Liu:2012ze,Mohler:2012na,Prelovsek:2014swa,Prelovsek:2013cra,Lee:2014uta,Lang:2015sba,Ikeda:2016zwx,%
Ikeda:2017mee,Cheung:2017tnt}.
To address these questions with confidence, it is important that lattice discretization (cutoff) effects be under control.
The more limited objective of the present work is to carry out a high-precision study of the
splittings of the low-lying charmonium states--particularly the $1S$ and $1P$ states--and, thus, lay the foundation for further
calculations of excited states.
Spin-dependent mass splittings are expected to be extremely sensitive to the charm-quark mass and to heavy-quark discretization
effects.
Reproducing these delicate splittings can therefore serve as another demonstration that systematic uncertainties are under excellent
control.

Our effort follows a previous analysis campaign on the same gauge configurations~\cite{Burch:2009az}.
Preliminary results have been reported~\cite{DeTar:2012xk} and some additional details about our quark sources can be found in
Ref.~\cite{Mohler:2014ksa}.
Our new results supersede the results in those publications.
Although other groups have reported partial results for the low-lying charmonium
spectrum~\cite{Dudek:2007wv,Bali:2011dc,Briceno:2012wt,Yang:2014sea,Galloway:2014tta}, there are no systematic, high-precision
studies for any action.
Thus, to our knowledge, our campaign is the first that includes precise tuning of the charm-quark mass, precise determination of the
lattice scale, and a controlled extrapolation to physical light sea-quark masses and zero lattice spacing.
Our paper is organized as follows.
In Sec.~\ref{story}, we further describe the objectives of our current work, while we describe our lattice setup in detail in
Sec.~\ref{methodology}.
Section~\ref{results} shows our results for the splittings among low-lying charmonium states including a chiral and continuum
extrapolation of results and a full error budget.
We summarize our findings in Sec.~\ref{conclusions} where we also provide a brief outlook.

\section{Theoretical Background}
\label{story}

In this paper, our main objective is the QCD determination of the masses of the 1S and 1P states in the charmonium spectrum.
From the ground state masses in the quantum number channels corresponding to the 1S and 1P states, we calculate the hyperfine
splitting between the 1S triplet and singlet states
\begin{equation}
    \Delta M_\text{HF} = M_{J/\psi}-M_{\eta_c} ,
\end{equation}
the spin-average 1P-1S splitting
\begin{align}
    \Delta M_\text{1P-1S}    &= M_{\overline{\text{1P}}} - M_{\overline{\text{1S}}},\\
    M_{\overline{\text{1P}}} &= \oneninth(M_{\chi_{c0}}+3M_{\chi_{c1}}+5M_{\chi_{c2}}),\\
    M_{\overline{\text{1S}}} &=  \quarter(M_{\eta_c}+3M_{J/\psi}),
\end{align}
and the spin-orbit, tensor, and 1P hyperfine splittings among the P-wave states
\begin{align}
    \Delta M_\text{spin-orbit} &= \oneninth(5M_{\chi_{c2}}-3M_{\chi_{c1}}-2M_{\chi_{c0}}),\\
    \Delta M_\text{tensor}     &= \oneninth(3M_{\chi_{c1}}-M_{\chi_{c2}}-2M_{\chi_{c0}}), \\
    \Delta M_\text{1PHF}       &= M_{\overline{\text{1P}}} -M_{h_c}\label{eq:1PHF}.
\end{align}
It is these splittings, extrapolated to zero lattice spacing and
physical sea-quark masses, that we compare with their experimental values.

These combinations are of phenomenological interest in constructing the heavy quarkonium potential, since they correspond to
separate terms in the potential derived from the heavy-quark limit~\cite{Eichten:1980mw,Peskin:1983up}.
\begin{align}
V_\text{tot} &= V(r) + V_S(r) \bm{S}_Q\cdot\bm{S}_Q + V_T(r) S_{12} + V_{LS}(r) \bm{L}\cdot\bm{S_Q}, \\
    S_{12} &= 3(\bm{S}_Q \cdot \hat{\bm{r}})(\bm{S}_Q \cdot \hat{\bm{r}}) - \bm{S}_Q \cdot \bm{S}_Q.
\end{align}
Thus, their dependence on the lattice spacing provides useful information
about discretization effects in each of the relevant terms, as discussed in detail in Ref.~\cite{Burch:2009az}.

\begin{table}[tbp]
\centering
\caption{Experimental averages of the masses and widths of the 1S and 1P low-lying charmonium states~\cite{Tanabashi:2018oca}.
}
\label{lowstates}
\begin{tabular}{ccc}
\hline
\hline
 \T\B meson & mass [MeV] & width \\
\hline
\T\B $\eta_c$ & 2983.9(5) & 32.0(8)~MeV\\
\T\B $J/\psi$ & 3096.900(6)& 92.9(2.8)~keV\\
\T\B $\chi_{c0}$ & 3414.71(30) & 10.8(6)~MeV\\
\T\B $\chi_{c1}$ & 3510.67(5) & 0.84(4)~MeV\\
\T\B $\chi_{c2}$ & 3556.17(7) & 1.97(9)~MeV\\
\T\B $h_c$ & 3525.38(11)& 0.7(4)~MeV\\
\hline
\hline
\end{tabular}
\end{table}

Table~\ref{lowstates} lists the 1S and 1P states along with their masses and widths, as determined from
experiment~\cite{Tanabashi:2018oca}.
While some of these states are extremely narrow, both the $\eta_c$ and $\chi_{c0}$ have a non-negligible hadronic decay width,
resulting from charm-anticharm annihilation.
In lattice QCD, this effect comes from disconnected diagrams, which our current simulation omits.
That is, we treat all low-lying charmonium states as stable.
It is therefore not \emph{a priori} clear whether we will obtain good agreement with the $\eta_c$ and $\chi_{c0}$ masses.
This shortcoming complicates the comparison with experiment, in particular for the 1S hyperfine
splitting.%
\footnote{Historically, the asymmetric line shape of the $\eta_c$ resonance also complicated the
extraction of the hyperfine splitting from experiment data.
This issue no longer arises with modern, high-statistics data (see,
for example, Refs.~\cite{BESIII:2011ab,Aaij:2016kxn}).}
We further comment on this issue when comparing our results with previous results in Sec.~\ref{comp_prev}.

\section{Methodology}
\label{methodology}

This section presents the methodology for the lattice determination of the charmonium masses described in Sec.~\ref{results}.
In addition to our lattice setup, our procedures to deal with uncertainties from the mistuning of the charm-quark mass, our strategy
for the chiral-continuum fits, and the systematic uncertainty arising from the determination of the lattice spacing are discussed in
this section.

\subsection{Gauge configurations}

\begin{table*}[tbp]
\centering
\caption{Parameters of the MILC ensembles used in this study.
    Listed are the lattice spacing~$a$, the ratio of the sea-quark masses $m_l^\prime/m_s^\prime$ used in the simulation, and the
    lattice size $L^3\times T$, Also included are the number of source time slices used in the calculation~$N_\text{src}$,
    the tuned charm-quark hopping parameter $\kappa_c$, the charm-quark hopping parameter of our simulation, $\kappa'_c$,
    and a citation for the ensemble.
    The first uncertainty in $\kappa_c$ is statistical, and the second is from the uncertainty in the lattice scale.}
\label{ensembles}
\begin{tabular}{S[table-format=1.3]cccccc}
\hline
\hline
 \T\B $\approx a$ [fm] & $m_l^\prime/m^\prime_s$ & size & $N_\text{src}$ & $\kappa_c$ & $\kappa'_c$ & Ref. \\
\hline
\T\B 0.14 & 0.2 & $16^3\times 48$ & 2524 & 0.12237(26)(20) & 0.1221 & \cite{asqtad:en00a} \\
\T\B 0.14 & 0.1 & $20^3\times 48$ & 2416 & 0.12231(26)(20) & 0.1221 & \cite{asqtad:en25a} \\
\T\B 0.114 & 0.2 & $20^3\times 64$ & 4800 & 0.12423(15)(16) & 0.12423 & \cite{asqtad:en06a,asqtad:en06b} \\
\T\B 0.114 & 0.1 & $24^3\times 64$ & 3328 & 0.12423(15)(16) & 0.1220, 0.1245, 0.1280 & \cite{asqtad:en04a} \\
\T\B 0.082 & 0.2 & $28^3\times 96$ & 1904 & 0.12722(9)(14) & 0.12722 & \cite{asqtad:en15a,asqtad:en15b} \\
\T\B 0.082 & 0.1 & $40^3\times 96$ & 4060 & 0.12714(9)(14) & 0.12714 & \cite{asqtad:en13a,asqtad:en13b} \\
\T\B 0.058 & 0.2 & $48^3\times 144$ & 2604 & 0.12960(4)(11) & 0.1298 & \cite{asqtad:en20a,asqtad:en20b} \\
\T\B 0.058 & 0.1 & $64^3\times 144$ & 1984 & 0.12955(4)(11) & 0.1296 & \cite{asqtad:en18a,asqtad:en18b} \\
\T\B 0.043 & 0.2 & $64^3\times 192$ & 3204 & 0.130921(16)(70)& 0.1310  & \cite{asqtad:en24a} \\
\hline
\hline
\end{tabular}
\end{table*}

We use the (2+1)-flavor gauge configurations generated by the MILC collaboration~\cite{Bazavov:2009bb} with the asqtad fermion
action for sea quarks.
The ensembles used in this work are listed in Table~\ref{ensembles}.
The use of five different lattice spacings $a$ and two different light sea-quark masses (given in the table as a fraction of the
strange quark mass in the simulation) enables us to perform a controlled chiral-continuum extrapolation.
Four source time slices per gauge configuration are used, for a total of approximately $2000$ to
$4000$ sources per ensemble.
We use the Fermilab prescription~\cite{ElKhadra:1996mp} for the charm quarks, which suppresses heavy-quark discretization effects in
mass splittings~\cite{Kronfeld:2000ck}.
The charm-quark hopping parameter $\kappa_c$ has been tuned by demanding that the $D_s$ kinetic mass be equal to the physical $D_s$
meson mass in the way described in Ref.~\cite{Bailey:2014tva}.
The resulting $\kappa_c$ and the (sometimes slightly different) simulation value $\kappa'_c$ are also
given in Table~\ref{ensembles}.
Note that we refer to the quark masses used in the simulation as $m_l^\prime$ and $m^\prime_s$ while we
denote the physical light- and strange-quark masses by $m_l$ and $m_s$.
When calculating observables, we need to take into account this difference in our chiral-continuum extrapolations.

\subsection{Calculation of observables\label{observables}}

We calculate a matrix of correlators $C(t)$ using quark-antiquark interpolators with the $J^{PC}$ quantum numbers of the states in
question, where $J$ is the total spin and $P$ and $C$ are parity and charge conjugation quantum numbers.
We opt for a basis built from interpolators with derivatives and use interpolating operators similar to those suggested by Liao and
Manke~\cite{Liao:2002rj}, which have also been used by Dudek \emph{et al.}~\cite{Dudek:2007wv}.
A subset of similar interpolators has also been used in Ref.~\cite{Gattringer:2008be} and similar interpolators using displacements
or full plaquettes rather than derivatives have previously been considered in Ref.~\cite{Lacock:1996vy}.
Disconnected contributions, where a valence charm-anticharm-quark pair annihilates, are omitted when calculating the correlators.

Our operators are constructed from stochastic wall sources, including covariant Gaussian smearing.
Stochastic sources consist of a four-component-spinor field on a single time slice with random color orientation, 
but definite spin:
\begin{equation}
  S^r_\beta(\bm{x},a,\alpha) = \eta^r_a(\bm{x}) \delta_{\alpha\beta},
\end{equation}
where $r$ labels the stochastic source, $\beta$ its spin, and $a$ and $\alpha$ are the 12 Dirac color and spin components.
Averaged over a sufficiently large number of stochastic sources $N_r$, we have
\begin{equation}
\lim_{N_r\rightarrow\infty} \frac{1}{N_r}\sum_{r=1}^{N_r} 
  \eta^{r*}_a(\bm{x}) \eta^{r}_b(\bm{y}) 
  = \delta_{ab} \delta_{\bm{x} \bm{y}}.
\end{equation}
With both charm and anticharm quarks originating from the same source, or with one source modified by Gaussian smearing, the
stochastic average gives the effect of charmonium sources composed of local or smeared bilinears of the form
\begin{equation}
    \mathcal{O}_i(x) = \bar{\psi}(x) O_i \psi(x),
    \label{eq:bilinears}
\end{equation}
where the smearing operators are included in the definition of $O_i$.
All links appearing in the Gaussian smearing operators and in the covariant derivatives below are smeared with a fixed number of
APE-smearing~\cite{Albanese:1987ds} steps.
Gaussian smearing is implemented by acting with a smearing operator $M$ on the stochastic sources $S$ to obtain Gaussian sources:
\begin{subequations}
\label{smearing_jlab}
\begin{align}
    G &= MS = \mathcal{N} \left(1 + \frac{\sigma^2}{4a^2N} \Delta\right)^NS ,
    \label{eq:GaussM} \\
    \Delta(\bm{x},\bm{y}) &= \sum_{i=1}^3\left[U_i(\bm{x},0)\delta(\bm{x}+a\bm{\hat{\imath}},\bm{y}) +
        U_i(\bm{x}-a\bm{\hat{\imath}},0)^\dagger\delta(\bm{x}-a\bm{\hat{\imath}},
        \bm{y})\right] -6\delta_{\bm{x}\bm{y}}, 
\end{align}
\end{subequations}
where $\Delta$ is a covariant 3D Laplacian, $\mathcal{N}$ is just a
normalization factor, and $\sigma/a$ and $N$ are chosen such that $M$ approximates a Gaussian with (physical) standard deviation $\sigma$ in coordinate space.
Thus,
\begin{equation}
    \lim_{N\to\infty} M = \mathrm{e}^{\sigma^2\Delta/4},
\end{equation}
because $\lim_{N\to\infty}(1+b/N)^N = e^b$.
Table~\ref{smearparms} lists the smearing parameters for both the gauge link smearing and for the Gaussian quark sources.

In the constructions discussed in Appendix~\ref{interpolators}, we use the following operators:
\begin{subequations}
    \label{operators}
\begin{align}
    \nabla_i     &= M P_iS, \label{eq:smearedNabla} \\
    \mathbb{B}_i &=  \varepsilon_{ijk} M P_jP_k S, \label{eq:smearedB} \\
    \mathbb{D}_i &= |\varepsilon_{ijk}|M P_jP_k S. \label{eq:smearedD}
\end{align}
\end{subequations}
Here $M$ is the Gaussian smearing operator defined in Eq.~(\ref{eq:GaussM}),
and $P_i$ is a derivative-type operator on a given time slice $t$,
\begin{align}
    P_i(\bm{x},\bm{y})&=\frac{1}{2}\left[W_i(\bm{x},t; \bm{x}+r\bm{\hat{\imath}}, t) \delta(\bm{x}+r\bm{\hat{\imath}},\bm{y}) -
        W_i(\bm{x}-r\bm{\hat{\imath}}, t;\bm{x},t) \delta(\bm{x}-r\bm{\hat{\imath}},\bm{y}) \right] ,
\end{align}
where $r$ is kept of roughly the same length in physical units and
$W_i(\bm{x},t; \bm{x}+r\bm{\hat{\imath}}, t)$ denotes the shortest Wilson
line connecting $(\bm{x},t)$ and $(\bm{x}+r\bm{\hat{\imath}}, t)$.
In Eqs.~(\ref{operators}), the~$P_i$ act to the right.
The continuum version of operator $\mathbb{B}_i$ has a relation to the chromomagnetic parts of the field strength tensor
\begin{align}
    \mathbb{B}_i^\text{cont}&=-\frac{i}{2}\varepsilon_{ijk}F^{jk} .
\end{align}
To avoid an (anti)symmetrization of the derivatives, which would require more sources, we first apply derivatives and then the Gaussian smearing.
A detailed discussion of this approach can be found in Ref.~\cite{Mohler:2009}.

\begin{table}[tbp]
    \centering
    \caption{Table of gauge link and quark smearing parameters.
    For the gauge link smearing $N_\text{APE}$ steps of APE smearing with smearing parameter $c$~\cite{Albanese:1987ds} have been
    applied.
    For the quark smearing detailed above, the standard deviation $\sigma$ is kept fixed at roughly $0.31~\text{fm}$ while the
    number of smearing steps $N$ is chosen suitably.}.
    \label{smearparms}
\begin{tabular}{S[table-format=1.3]ccccc}
\hline
\hline
 \T\B $\approx a$ [fm]& $m_l^\prime/m_s^\prime$ & $N_\text{APE}$ & $c$ & $\sigma/a$ & $N$\\
\hline
\T\B 0.14  & 0.2 & 15 & 0.1 & 2.2 & 20\\
\T\B 0.14  & 0.1 & 15 & 0.1 & 2.2 & 20\\
\T\B 0.114 & 0.2 & 15 & 0.1 & 2.6 & 40\\
\T\B 0.114 & 0.1 & 15 & 0.1 & 2.8 & 20\\
\T\B 0.082 & 0.2 & 15 & 0.1 & 3.7 & 20\\
\T\B 0.082 & 0.1 & 15 & 0.1 & 3.7 & 50\\
\T\B 0.058 & 0.2 & 15 & 0.1 & 5.5 & 80\\
\T\B 0.058 & 0.1 & 15 & 0.1 & 5.5 & 80\\
\T\B 0.043 & 0.2 & 15 & 0.1 & 7.0 &100\\
\hline
\hline
\end{tabular}
\end{table}

We use the variational method~\cite{Michael:1985ne,Kronfeld:1989tb,Luscher:1990ck,Blossier:2009kd}, solving the
generalized eigenvalue problem
\begin{align}
C(t)\vec{\psi}^{(k)}&=\lambda^{(k)}(t)C(t_0)\vec{\psi}^{(k)},\\
\lambda^{(k)}(t)&\propto\mathrm{e}^{-tE_k}\left(1+\order\left(\mathrm{e}^{-t\Delta E_k}\right)\right),
\end{align}
with reference time slice $t_0$.
The ground state mass can be extracted from the large time behavior of the largest eigenvalue.
For this we use (multi)exponential fits to the eigenvalues in the interval $[t_\text{min},t_\text{max}]$, taking into account
correlations in time separation.
At fixed $t_0$, $\Delta E_k$ is formally given by
\begin{align}
\Delta E_k &= \mathrm{min}|E_m-E_n|,\qquad m\neq n  ,
\end{align}
while for the special case of $t\le 2t_0$ and a basis of $N$
correlators~\cite{Blossier:2008tx} $\Delta E_k$ is given by
\begin{align}
\Delta E_k &= E_{N+1}-E_n  .
\end{align}
We investigate the dependence of our results on $t_0$ and find that in practice a rather small value of $t_0$ provides the best
compromise between excited-state contaminations and statistical uncertainty.
Here and elsewhere, the statistical uncertainties are computed from a single-elimination jackknife.
In our analysis, the reference time $t_0$ and the lower boundary of the fit window $t_\text{min}$ are kept roughly constant in fm 
for the 1S and 1P states respectively.%
\footnote{For one of the ensembles at lattice spacing $a=0.082$~fm, one of the multiexponential fits is not stable 
with our usual value of $t_\text{min}$, so we choose a smaller $t_\text{min}=0.25~\text{fm}$.
We stress that the results on this ensemble are fully compatible with single-exponential fits at large time separations, and that
our final results are not affected by this choice.} %
The upper boundary of the fit-window $t_\text{max}$ is chosen such that the eigenvectors $\vec\psi^{(k)}$ remain stable within
statistics in the whole fit range, which in some cases results in a somewhat shorter fit windows than just considering plateaus in
the effective masses.
For the P-wave states on the coarsest lattice spacing, where the tuning of the quark-smearing was performed, remaining excited-state
contaminations are extremely small, and we need to use loose priors on the mass splittings between the ground state and the lowest
excitations in order to avoid clearly unphysical fit results with two almost mass-degenerate ground states.

In some cases increasing the size of the basis used in the variational method leads to no improvement in the ground state but adds
statistical noise.
For our final results we therefore opted to suitably prune the interpolator basis, and we list our choices of basis in
Appendix~\ref{interpolators}.

\subsection{Charm-quark-mass corrections}
\label{sec:correction}

\begin{table}[tbp]
\centering
\caption{Values used to correct for charm-quark-mass mistunings for each of the ensembles in this study.
  Shown are the approximate ensemble lattice spacing, the ratio of simulation sea-quark masses
  the critical $\kappa$ value, the tadpole factor, and the factor $A$ from Eq.~(\ref{eq:A}).}
\label{avalues}
\begin{tabular}{S[table-format=1.3]cS[table-format=1.6]S[table-format=1.5]S[table-format=3.2]}
\hline
\hline
 \T\B $\approx a$ [fm] & $m_l^\prime/m_s^\prime$ & {$\crit$} & {$u_0$} & {$A$} \\
\hline
\T\B 0.14  & 0.2 & 0.142432 & 0.8604 & 71.54\\
\T\B 0.14  & 0.1 & 0.14236  & 0.8602 & 71.15\\
\T\B 0.114 & 0.2 & 0.14091  & 0.8677 & 85.06\\
\T\B 0.114 & 0.1 & 0.14096  & 0.8678 & 85.04\\
\T\B 0.082 & 0.2 & 0.139119 & 0.8782 & 112.42\\
\T\B 0.082 & 0.1 & 0.139173 & 0.8779 & 111.51\\
\T\B 0.058 & 0.2 & 0.137632 & 0.88788& 155.40\\
\T\B 0.058 & 0.1 & 0.137678 & 0.88764& 154.10\\
\T\B 0.043 & 0.2 & 0.13664  & 0.89511& 208.69\\
\hline
\hline
\end{tabular}
\end{table}

For some of the ensembles listed in Table~\ref{ensembles} the charm-quark hopping parameter of the simulation
$\kappa'_c$ differ slightly from the physical charm-quark hopping parameter $\kappa_c$.
The raw splittings on these ensembles have to be corrected for this mistuning.
To determine the needed correction, we compute the derivative of each mass splitting with respect to $\kappa_c$ on one ensemble,
namely the one with $a=0.114$~fm and $m_l/m_h=0.1$~\cite{asqtad:en04a}.
We assume that once the slope is expressed in terms of physical quantities, it remains the same for that mass splitting for all
ensembles.
Since the adjustments are small, any residual lattice spacing dependence in the slopes should be negligible.

To be explicit, for mass splitting $\Delta M_i$, we assume that the following derivative is the same for all ensembles:
\begin{equation}
  R_i = \frac{d\Delta M_i}{dm_2} ,
\end{equation}
where $m_2(\kappa_c)$ is the kinetic mass of the charm quark.
For a given $\kappa_c$ we estimate that mass from the ensemble's critical hopping parameter $\crit$ and tadpole factor $u_0$
using the tree-level expressions [Eq.~(4.9) of~\cite{ElKhadra:1996mp}]:
\begin{align}
     am_0 &= \frac{1}{2 u_0} \left(\frac{1}{\kappa_c} - \frac{1}{\crit}\right), \\
     \frac{1}{am_2} &= \frac{2}{am_0(2+am_0)} + \frac{1}{1+am_0} .
\end{align}
The correction to the mass splitting, resulting from a shift
$d\kappa_c$ is then given in $r_1$ units~\cite{Bernard:2000gd} by
\begin{equation}
  r_1\, d\Delta M_i = R_i A\, d\kappa_c ,
\end{equation}
where
\begin{equation}
  A  = \frac{dam_2}{d\kappa_c}\frac{r_1}{a} .
\label{eq:A}
\end{equation}
Values of $\crit$, $u_0$, and $A$ for each ensemble are listed in Table~\ref{avalues}.
A quantitative estimate for the uncertainty from this procedure is provided in Sec.~\ref{uncertainty}.

\subsection{Chiral and continuum fits}

\begin{table}[tbp]
\centering
\caption{Simulation light and heavy sea-quark masses compared with physical light and strange quark 
masses for each ensemble.}
\label{quarkmasses}
\begin{tabular}{S[table-format=1.3]S[table-format=1.4]S[table-format=1.4]ccc}
\hline
\hline
 \T\B $\approx a$ [fm] & $am_l^\prime$ & $am_s^\prime$ & $am_l$ & $am_s$ & $\alpha_s(2/a)$ \\
\hline
\T\B 0.14  & 0.0097 & 0.0484 &  0.0015079 & 0.04185 & 0.35885 \\
\T\B 0.14  & 0.0048 & 0.0484 &  0.0015180 & 0.04213 & 0.36042 \\
\T\B 0.114 & 0.01   & 0.05   &  0.0012150 & 0.03357 & 0.31054 \\
\T\B 0.114 & 0.005  & 0.05   &  0.0012150 & 0.03357 & 0.31035 \\
\T\B 0.082 & 0.0062 & 0.031  &  0.0008923 & 0.02446 & 0.26062 \\
\T\B 0.082 & 0.0031 & 0.031  &  0.0009004 & 0.02468 & 0.26177 \\
\T\B 0.058 & 0.0036 & 0.018  &  0.0006401 & 0.01751 & 0.22451 \\
\T\B 0.058 & 0.0018 & 0.018  &  0.0006456 & 0.01766 & 0.22531 \\
\T\B 0.043 & 0.0024 & 0.014  &  0.0004742 & 0.01298 & 0.20131 \\
\hline
\hline
\end{tabular}
\end{table}

\begin{figure}[tbp]
\centering
\includegraphics[clip,height=6cm]{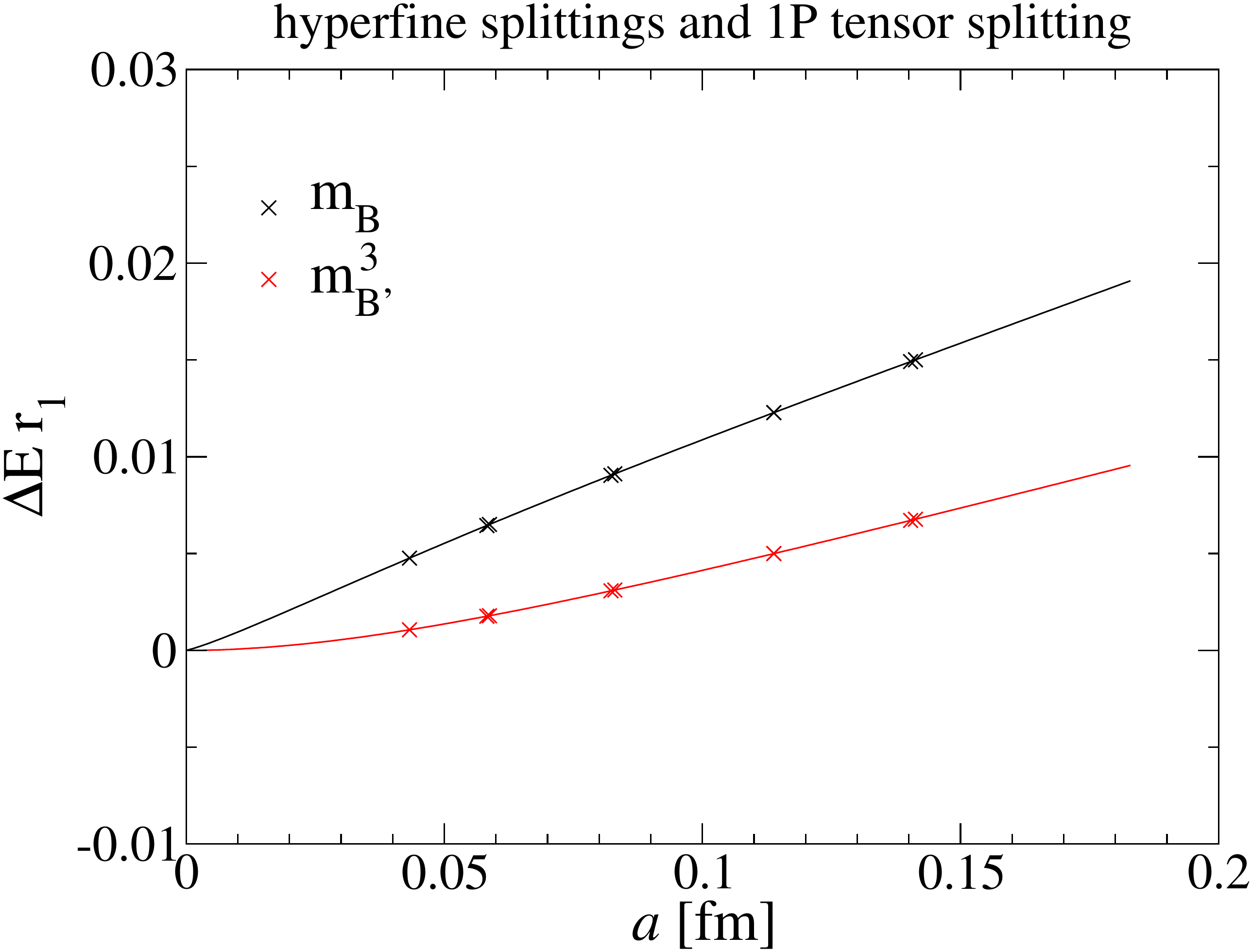}
\includegraphics[clip,height=6cm]{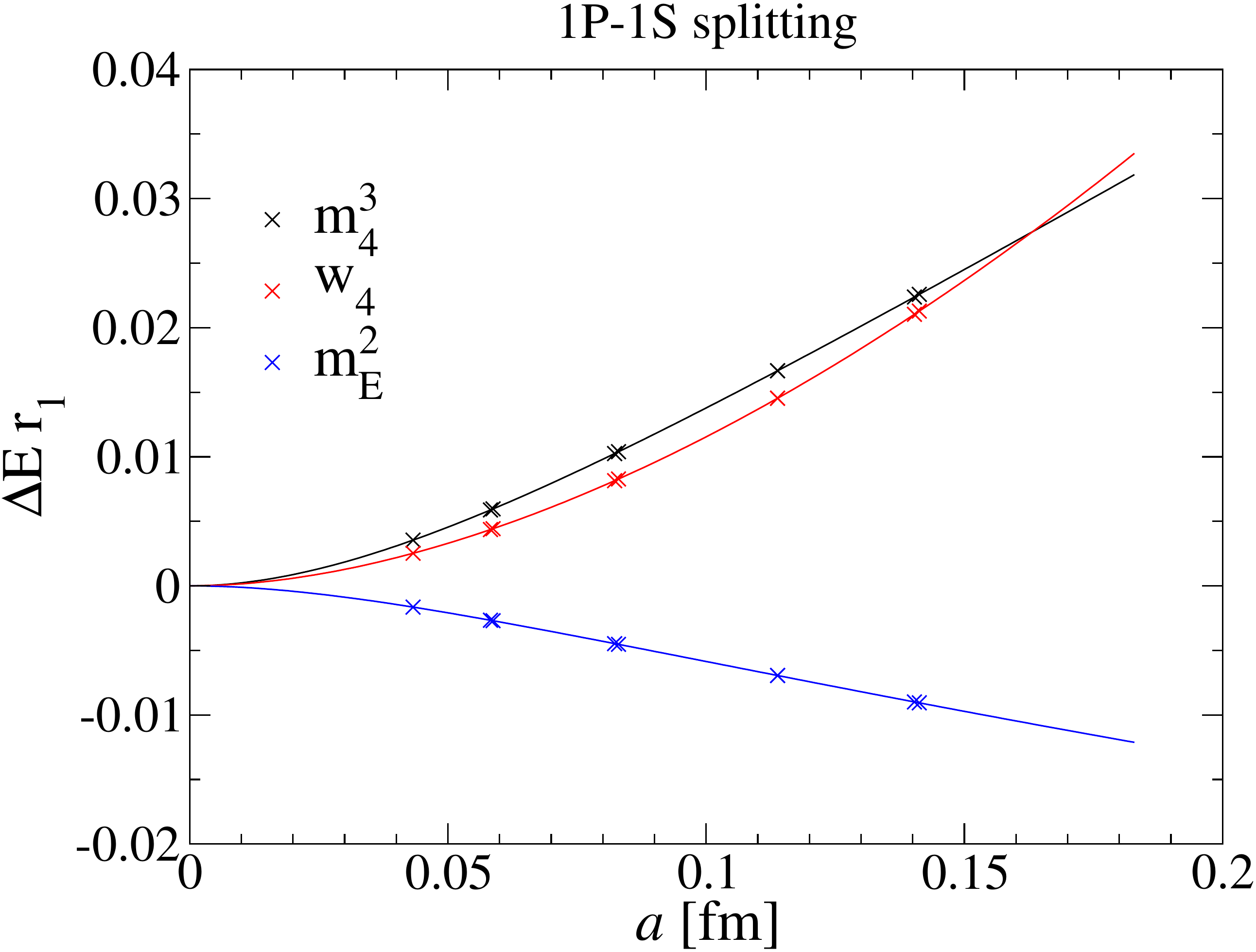} \\
\includegraphics[clip,height=6cm]{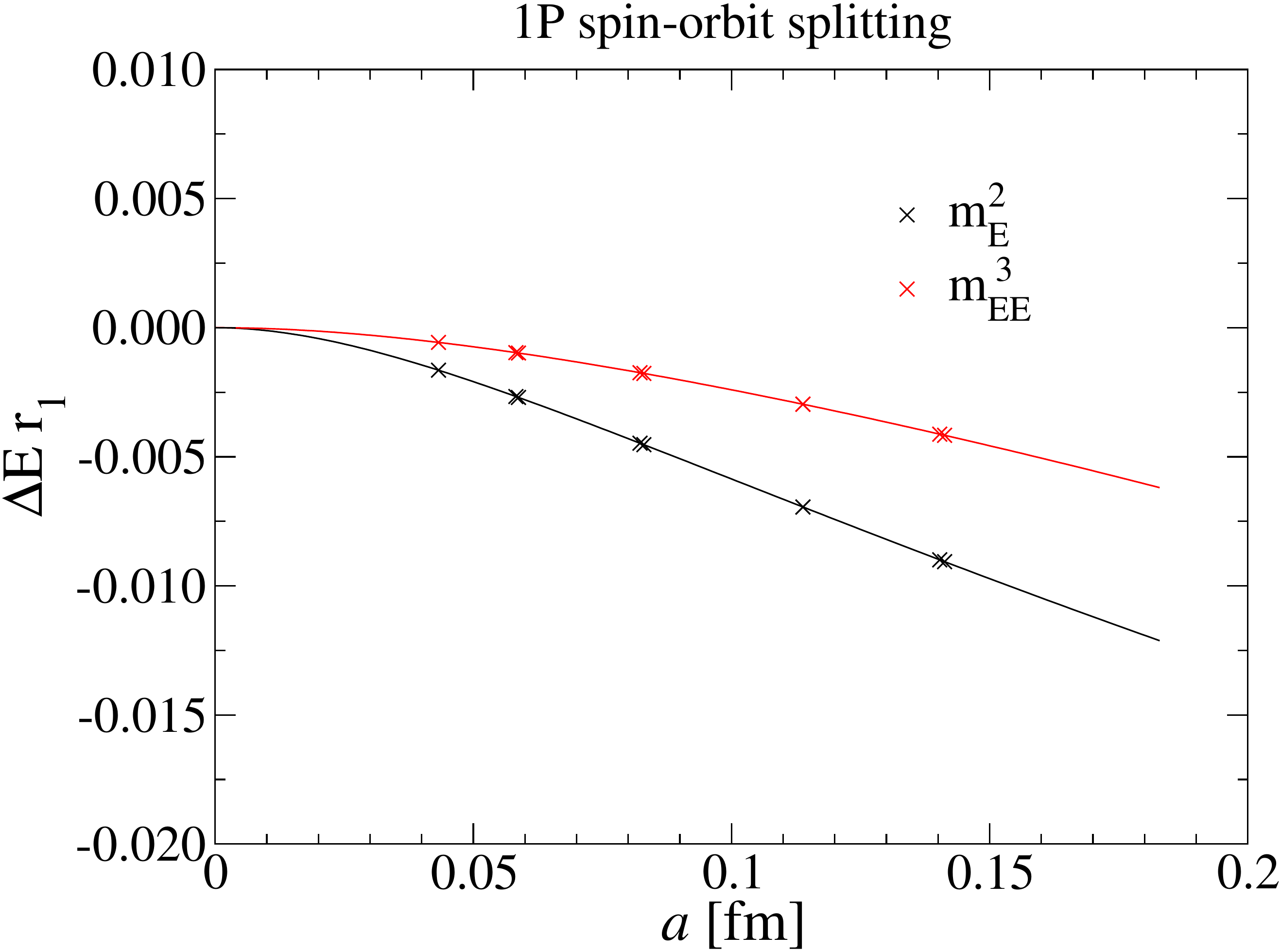}
\caption{Shapes and size of the expected heavy-quark discretization
  uncertainties for charmonium splittings (NRQCD power counting~\cite{Lepage:1992tx}) in the
  Fermilab approach (using $v^2=0.3$ and $mv^2\approx 420~\text{MeV}\approx$
  1P-1S splitting). These are as in Figs.~3 and 4 of Ref.~\cite{Oktay:2008ex},
  and the notation for the terms follows that reference.
  Values of $\alpha_s$ consistent with those in Table~\ref{quarkmasses} have
  been used. The terms arising from mass mismatches are denoted in the plot by the masses in the
  short-distance coefficients. In addition a rotational symmetry breaking term
  (with coefficient $w_4$) is important for the 1P-1S splitting. Expressions
  for the short-distance coefficients can be found in Ref.~\cite{Oktay:2008ex}.
}
\label{shapes}
\end{figure}

We perform a combined extrapolation to the continuum values and to physical light- and strange-quark masses.
Our data indicate a clear sea-quark mass dependence for some of the observables,%
\footnote{This effect is particularly noticeable in the sea-quark-mass-independent renormalization scheme~\cite{Bazavov:2009bb},
which we have adopted here.} which means that we also need to take into account the effect of mistuned strange sea-quark masses.
Our model for the lattice spacing dependence is based on the Oktay-Kronfeld~\cite{Oktay:2008ex} analysis of the Fermilab
prescription, which provides NRQCD power-counting~\cite{Lepage:1992tx} estimates of various heavy-quark discretization effects in
quarkonium.
They are parameterized as mass mismatches, leading to functions $f_i(a)$ of the lattice spacing that are determined separately for
each observable.
In addition to the terms for the heavy-quark discretization effects, we also add a term linear in $\alpha_s a^2$ as appropriate for
the asqtad sea quarks.
For our combined sea-quark mass and continuum fit we use the Ansatz
\begin{align}
    M  &= M_0+b(2x_l+x_s)+c_0f_1(a,\alpha_s)+c_1f_2(a,\alpha_s)+\cdots, \label{eq:Ansatz} \\
    x_l&= \frac{m_l^\prime-m_l}{m_s}, \\
    x_s&= \frac{m_s^\prime-m_s}{m_s}
\end{align}
as our fit model.
The values for $m_l^\prime$, $m_l$, $m_s^\prime$ and $m_s$ are given in Table~\ref{quarkmasses} along with the values of the
renormalized coupling in the $V$~scheme~\cite{Lepage:1992xa} $\alpha_s$ at scale $2/a$ used in the analysis of discretization
effects.
For each observable we determine the most important mass mismatches arising at $\order(v^4)$ and/or $\order(v^6)$ in NRQCD
power-counting.
Figure~\ref{shapes} shows the expected discretization uncertainties from power counting estimates for the splitting indicated in the
respective figure.
The plotted curves correspond to $c_i=1$.
For the 1P-1S splitting, this includes a term from rotational symmetry breaking ($w_4$ term).
In our default fits we use Bayesian priors centered around 0 with a prior uncertainty of 1 as a constraint for all terms originating
from heavy-quark discretization effects.
As part of our systematic variations described in Sec.~\ref{uncertainty}, this prior uncertainty is varied.
In addition to these terms we also allow for a generic $\alpha_sa^2$ term
(without prior) characteristic of
light-quark discretization effects.
We discuss the relevant mass mismatches for a given splitting when we present our results in Sec.~\ref{results}.
For each observable we compare continuum extrapolations with just two terms ($\alpha_sa^2$ and the leading heavy-quark
discretization term) and with three terms (the $\alpha_sa^2$ term and the leading {\em and} subleading heavy-quark discretization
terms).
We further check the variation from replacing the $\alpha_s a^2$ term by an $a^2$ term.
While a single leading shape is usually enough to get a good fit of the data, including further possible shapes leads to a larger
and more realistic uncertainty estimate.
The fit variations described above are among the fit variations shown in Sec.~\ref{uncertainty}, where our error budget is also
discussed.

\subsection{Scale-setting uncertainty}
\label{scale_error}

For the figures presented in Sec.~\ref{results}, we use MILC's version of the Sommer scale, $r_1$~\cite{Bernard:2000gd}.
The values of $r_1/a$ for the asqtad ensembles and an explanation for our value $r_1 = 0.31174(216)$~fm can be found in
Ref.~\cite{Bazavov:2011aa}.
This value was determined in the ``mass-independent'' scale-setting scheme, the one adopted here.
To estimate the scale-setting error, for each observable we first determine the result using the central value for both $r_1$ and
$\kappa_c$ and then repeat the procedure, shifting $r_1$ by one standard deviation while simultaneously shifting the tuned
$\kappa_c$ by an amount that results from the same shift in $r_1$.
The scale-setting uncertainty for each observable is discussed in Sec.~\ref{uncertainty} and tabulated in Table~\ref{errorbudget}.

\section{Results}
\label{results}

In this section, results for the mass splittings from Sec.~\ref{story} are
presented. After discussing each splitting in turn, the systematic
uncertainties associated with the determination are quantified and the
resulting values are compared with the results from previous determinations.

\subsection{1S hyperfine splitting}

\begin{figure}[tbp]
\centering
\includegraphics[clip,width=10.0cm]{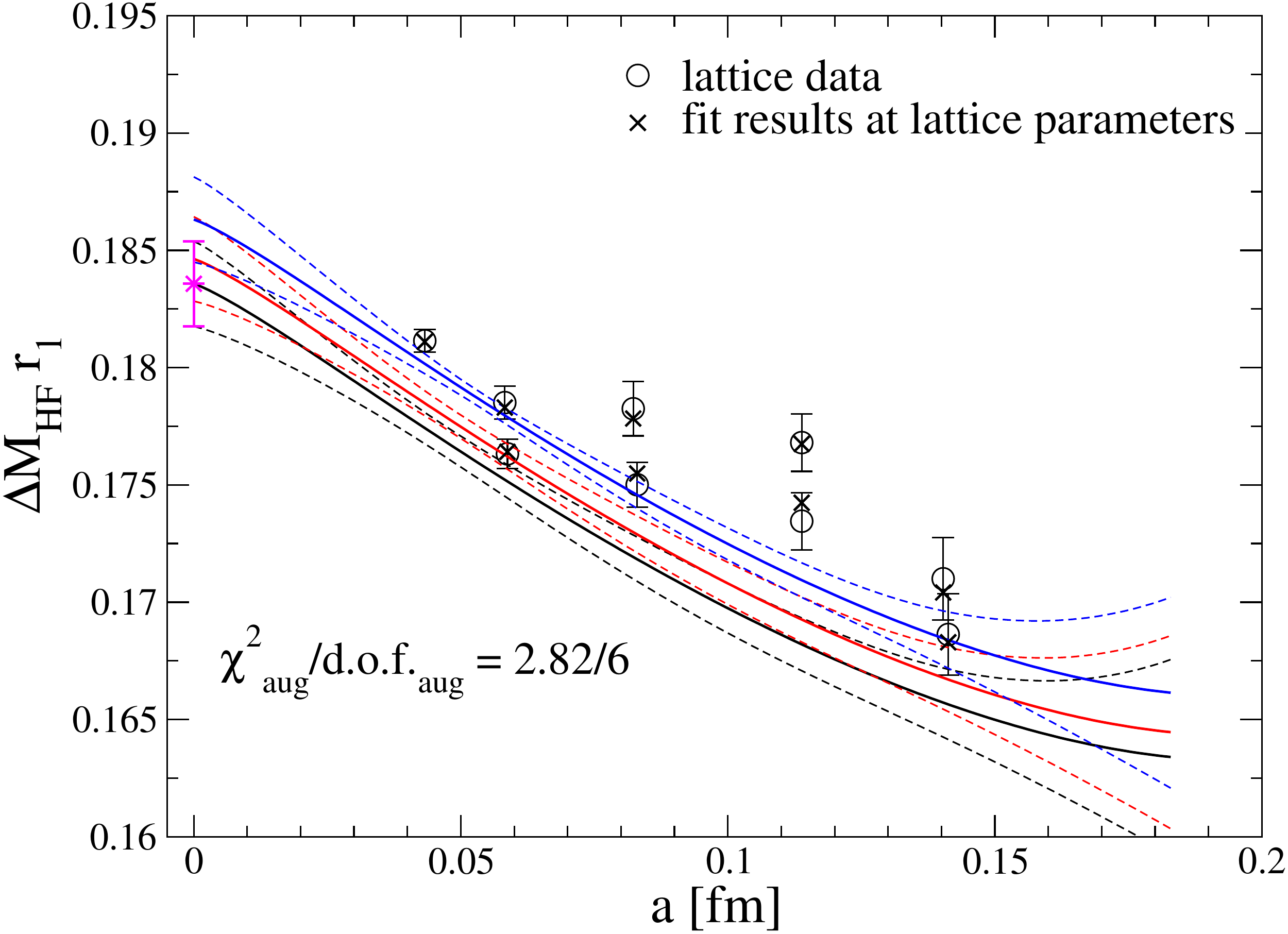}
\caption{Chiral and continuum fit for the 1S hyperfine splitting.
The black circles denote the lattice data. Curves for physical (black),
$0.1m_s$, (red) and $0.2m_s$ (blue) light-quark masses are plotted. Due to the
mistuning of the strange quark in the sea, which differs from ensemble to
ensemble, the data points appear away from the curves. To illustrate that the
data are well described by the fit, the black crosses show the fit results
evaluated at the lattice parameters of the gauge ensemble. 
The magenta symbol indicates the result in the combined chiral and continuum limit.}
\label{1Shf}
\end{figure}

Like all spin-dependent splittings, the 1S-hyperfine splitting is highly sensitive to heavy-quark discretization and charm-quark
tuning effects. As such, it is an important benchmark quantity for lattice-QCD
calculations of the charmonium spectrum.

For the 1S hyperfine splitting, autocorrelations in the Markov chain of gauge configurations are significant and need to be taken
into account.
To do so, we estimate the integrated autocorrelation time using two methods.
Method one is to determine the autocorrelation time from the jackknife sample using the method and software of
Wolff~\cite{Wolff:2003sm}.
An alternative consists of constructing binned data from the jackknife estimates of the unbinned set and extrapolating the results
for bins of sizes 1 to 5 to infinite binsize using the expected scaling.
We determined the integrated autocorrelation time using both methods and check the results for consistency.
The two methods agree excellently and, as the uncertainty estimate on different ensembles is independent, we use the second method to 
inflate the statistical uncertainties on a single ensemble appropriately.%
\footnote{Note that the fit results in Table~\ref{swave_table} in the
Appendix~\ref{app:levels} are the uninflated results from the plateau fits.
The corresponding $\chi^2/\mathrm{d.o.f.}$ reflects the non-negligible autocorrelations.}

Figure~\ref{1Shf} shows the results for the 1S hyperfine splitting, along with a chiral-continuum extrapolation of the results.
Where needed, the data have already been shifted for mistuning of the charm-quark hopping parameter, as outlined in
Sec.~\ref{sec:correction}.
Note that significant contributions from charm-annihilation diagrams to this observable are expected~\cite{Levkova:2010ft}.
When comparing our final results with the experimental value in Table~\ref{numbers} we use the determination of $-1.5$ to $-4$~MeV
from Ref.~\cite{Levkova:2010ft} as an estimate for the uncertainty from neglecting disconnected contributions.

The leading heavy-quark discretization effects contributing to the hyperfine splittings come from mismatches of $m_B$ and $m_2$.
Following Ref.~\cite{Oktay:2008ex}, we use NRQCD power counting with $v^2=0.3$ and $m_c=1400$~MeV along with the tree level formulas
from Ref.~\cite{ElKhadra:1996mp} to estimate the expected size of all heavy-quark discretization effects.
The relevant formula for $m_B$ is Eq.~(4.22) of Ref.~\cite{Oktay:2008ex}, and the shape of the resulting mismatch is plotted in the
first pane of Fig.~\ref{shapes}.
Note that our fermion action includes a clover term~\cite{Sheikholeslami:1985ij} with the tadpole-improved tree-level value $c_B=c_E=u_0^{-3}$, where
$u_0$ is the average link from the plaquette.
This contribution is therefore suppressed relative to $m_cv^2$ (the kinetic
energy of the meson) by a factor $\frac{1}{2}\alpha_sv_c^2$. The sign of the contribution is, however, not known.

The next largest heavy-quark discretization effects come from mismatches of $m_{B^\prime}$ and $m_2$, where the relevant formula for
$m_{B^\prime}$ is given by Eq.~(4.23) of Ref.~\cite{Oktay:2008ex}.
Again, the resulting estimate of discretization effects from the mismatch is plotted in the first pane of Fig.~\ref{shapes}.
Note that at tree-level and with only terms up to dimension 5 in the action, this mismatch is the same as the one from the
difference between $m_{4}$ and $m_2$ (see below) but it is of a higher order in the NRQCD power counting and therefore suppressed by
$\frac{1}{8}v^4$ with respect to the kinetic energy.
For our final fits we use both of these mass mismatches with priors for the coefficients $c_i$ from Eq.~(\ref{eq:Ansatz}) given by
$0\pm 1$ as well as an unconstrained $\alpha_sa^2$ term.
The expected shapes for the mismatches are plotted in Fig.~\ref{shapes}.

\begin{table}[tbp]
\centering
\caption{Description of the variations in the chiral-continuum fit plotted in Fig.~\ref{fitvariations}.}
\label{fitvariations}
\begin{tabular}{cl}
\hline
\hline
A & same as ``default'' but using sea-quark discretization effects of  order $a^2$ rather than $\alpha_s a^2$\\
B & results when omitting the lattice data at the coarsest lattice spacing\\
C & results when omitting the lattice data at the finest lattice spacing\\
D & result using just terms of order $\alpha_s a^2$ and a single shape for the
heavy-quark \\ & discretization effects\\
E & heavy-quark discretization effects with priors for $c_i$ half of the default width ($0\pm 0.5$) \\
F & heavy-quark discretization effects with priors for $c_i$  double the default width ($0\pm 2$)   \\
G & $1\sigma$ variation of the $\kappa_c$ slope used to shift data to
physical $\kappa_c$\\
\hline
\hline
\end{tabular}
\end{table}

Finally, the stability of results with regard to systematic variations of the chiral-continuum fit needs to be assessed.
Table~\ref{fitvariations} describes a number of important fit variations (A--G), and their effect on the 1S hyperfine
splitting can be seen in the first pane of Fig.~\ref{chi_cont_var} in Sec.~\ref{uncertainty}, below.
One of these variations (D) consists of limiting the continuum extrapolations
to just two shapes (leading heavy-quark mismatch and sea-quark
term).
From the difference between the default value and D it can be seen that the central value is largely unaffected while the
uncertainty estimate from the fit with leading and subleading shapes is more conservative.
Note also that the fit results are stable when the default prior widths are doubled (variation F in
Table~\ref{fitvariations} and Fig.~\ref{chi_cont_var}).
While the results are stable when omitting the finest lattice spacing (variation C) there is a somewhat significant shift when
excluding the coarsest lattice spacing (variation B).
Therefore we take the difference between B and the default fit model as an additional systematic uncertainty.
For our final uncertainty estimate provided in Table~\ref{errorbudget} the uncertainties from the scale determination (direct and
through the uncertainty in the charm-quark mass) and from the correction of the data for simulation at unphysical charm-quark mass
are non-negligible.

\subsection{1P-1S splitting}

\begin{figure}[tbp]
\centering
\includegraphics[clip,width=10.0cm]{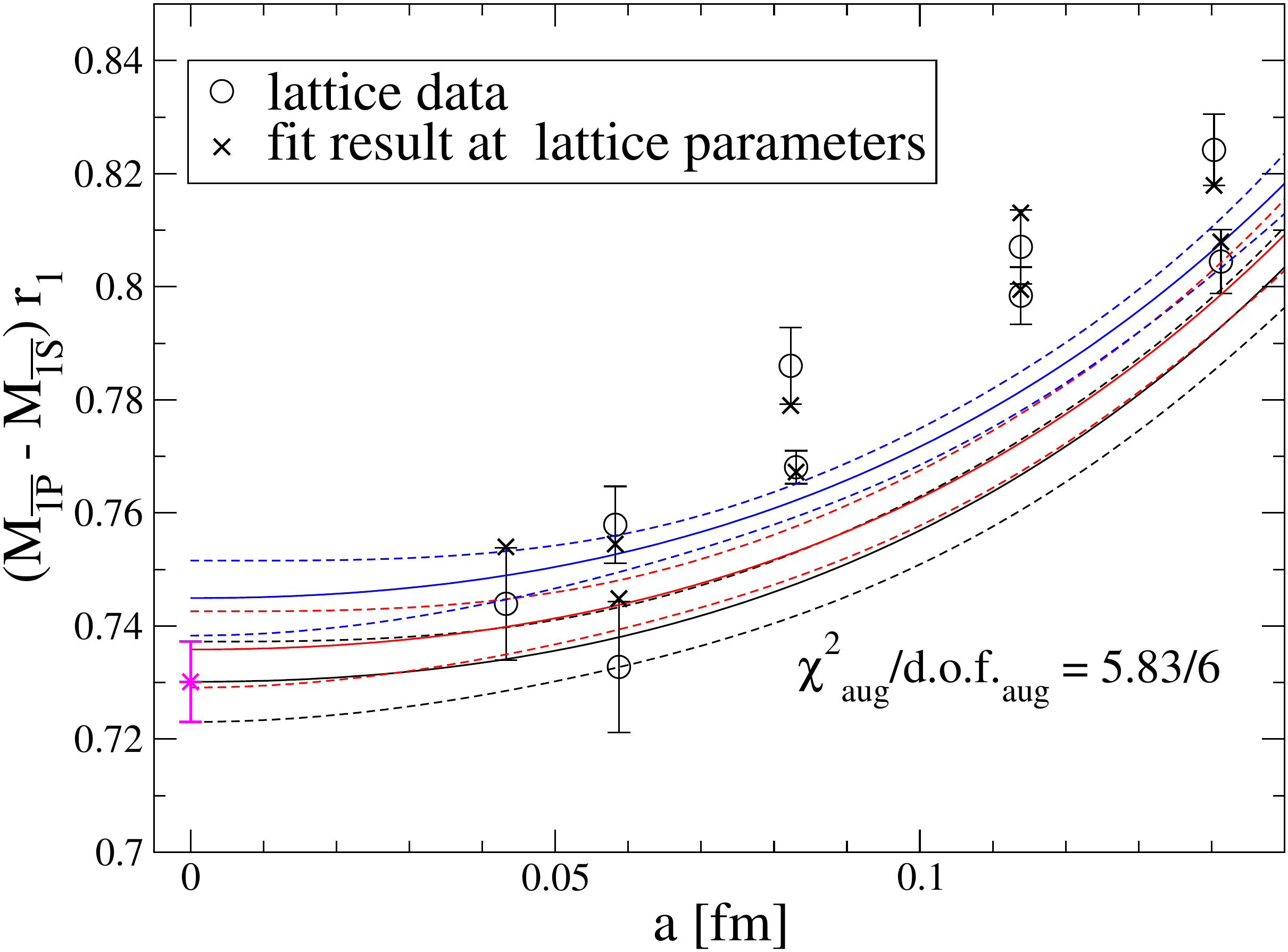}
\caption{Chiral and continuum fit for the 1P-1S splitting. The black circles
  denote the lattice data. Curves for physical (black),
  $0.1m_s$, (red) and $0.2m_s$ (blue) light-quark masses are plotted. The black
  crosses show the fit results evaluated at the lattice parameters of
  the gauge ensemble. The magenta symbol indicates the result in the combined
  chiral and continuum limit.}
\label{1P1S}
\end{figure}

Figure~\ref{1P1S} shows our result for the splitting between the spin-averaged P- and S-wave states $\Delta M_\text{1P-1S}$.
As in the 1S hyperfine splitting, significant effects from mistuned strange-quark masses are visible in our data.
Unlike the hyperfine splitting, there is no statistically significant autocorrelation in the Monte Carlo chain, and we therefore
treat the data as uncorrelated. In this case, we find large discretization
effects, emphasizing the need for several lattice spacings.

Having normalized the kinetic energy correctly, we expect leading heavy-quark discretization effects of order $v^4$ in NRQCD power
counting and we plot the expected shapes of the discretization effects in Fig.~\ref{shapes}.
The terms from the mismatch of $m_4$ and $m_2$ and the rotational symmetry breaking term arising at order $p^4$ are of about equal
size.
The relevant formulae for $w_4$ and $m_4$ are Eqs.~(4.4) and~(4.5) of Ref.~\cite{Oktay:2008ex}.
We also consider the discretization effects from the mismatch of $m_E$ and $m_2$, where $m_E$ is given by Eq.~(4.17) of
Ref.~\cite{Oktay:2008ex}, and we evaluate $m_E$ for $c_E=c_B=1$.
Again we use Bayesian priors with default value $0\pm1$ for the coefficients $c_i$ in Eq.~(\ref{eq:Ansatz}) associated with
heavy-quark discretization effects.

The chiral-continuum fits are stable under all variations shown in
Table~\ref{fitvariations}. The effect of these variations is illustrated in the second
pane of Fig.~\ref{chi_cont_var}.
In particular the central values do not change when the prior width is increased.
As for all other splittings, our final result based on the default fit model also takes into account possible discretization effects
of order $\alpha_s a^2$.
The largest variation with respect to this fit model occurs when replacing this term by an $a^2$ term, which is not motivated by the
sea quark action used.

\subsection{Spin-dependent P-wave splittings}

\begin{figure}[tbp]
\centering
\includegraphics[clip,width=10.0cm]{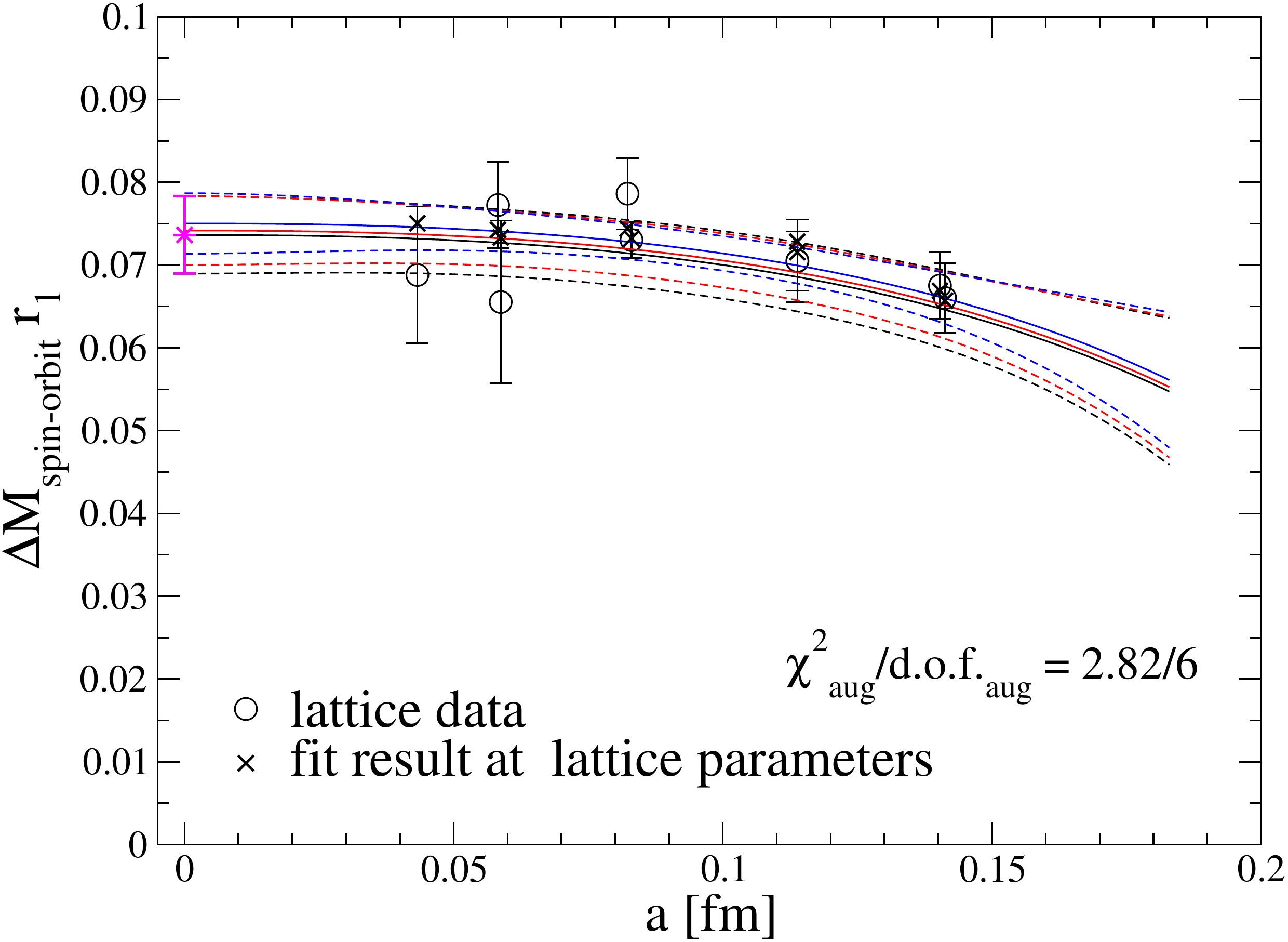}
\caption{Chiral and continuum fit for the 1P spin-orbit splitting. The black
  circles denote the lattice data. Curves for physical (black),
  $0.1m_s$, (red) and $0.2m_s$ (blue) light-quark masses are plotted. The black
  crosses show the fit results evaluated at the lattice parameters of
  the gauge ensemble. The magenta symbol indicates the result in the combined
  chiral and continuum limit.}
\label{1Pso}
\end{figure}

\begin{figure}[tbp]
\centering
\includegraphics[clip,width=10.0cm]{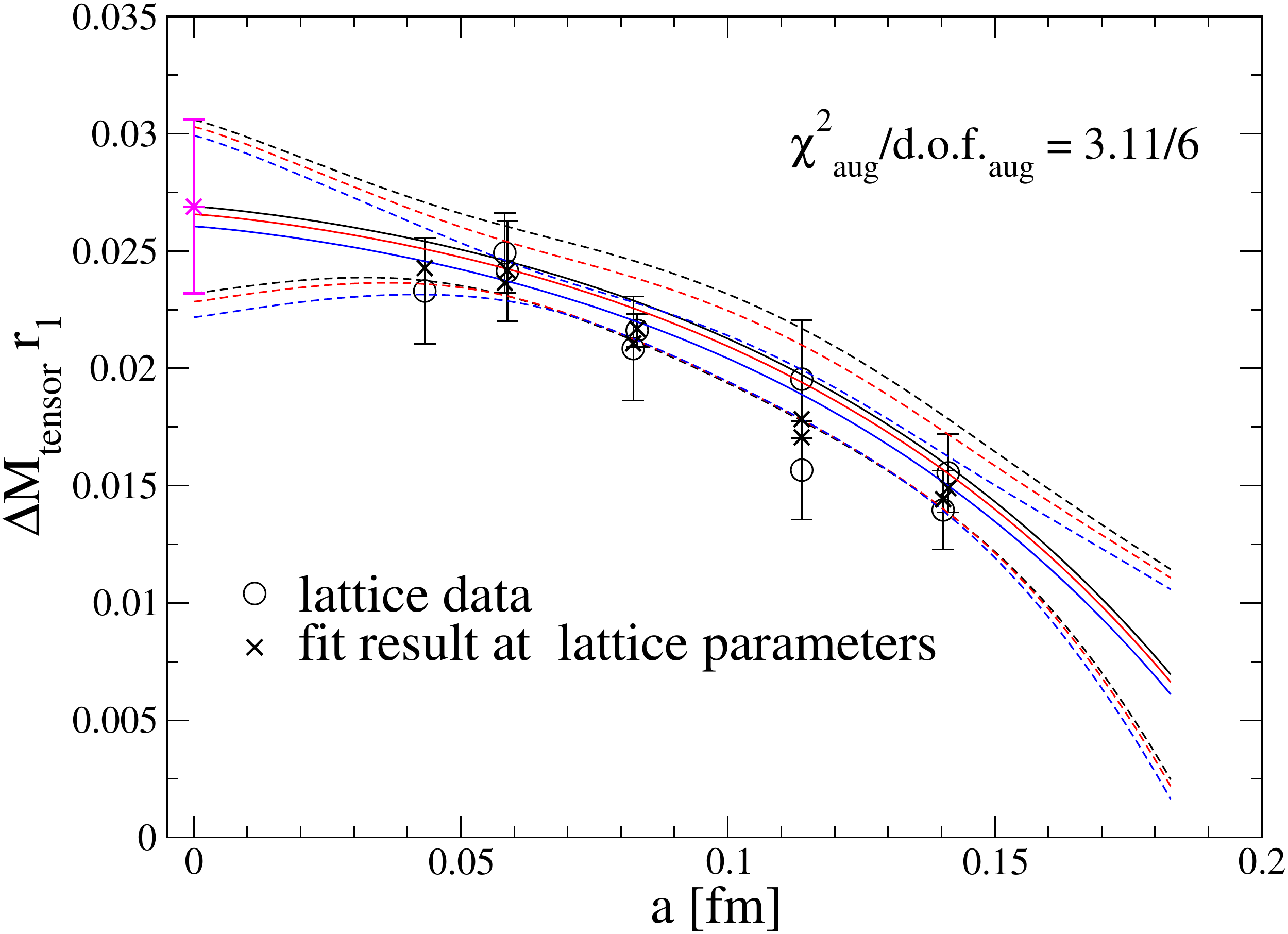}
\caption{Chiral and continuum fit for the 1P tensor splitting. The black circles denote the lattice data.
  Curves for physical (black),
  $0.1m_s$, (red) and $0.2m_s$ (blue) light-quark masses are plotted. The black
  crosses show the fit results evaluated at the lattice parameters of
  the gauge ensemble. The magenta symbol indicates the result in the combined
  chiral and continuum limit.}
\label{1Pte}
\end{figure}

The P-wave spin-orbit splitting---shown in Fig.~\ref{1Pso}---exhibits only small discretization uncertainties.
The leading heavy-quark discretization effects come from the mismatch of $m_E$ and $m_2$ and are of order $v^4$ in NRQCD power
counting.
Subleading effects of oder $v^6$ come from the mismatch of $m_{EE}$ and $m_2$ and from terms of mass-dimension eight not considered
in Ref.~\cite{Oktay:2008ex}.
Our default fit employs Bayesian priors given by $0\pm1$ for all relevant shapes from Fig.~\ref{shapes}.

The results for the spin-orbit splitting are very stable with respect to the variations of the chiral-continuum fit in
Table~\ref{fitvariations}.
The results of this variation are shown in the third pane of Fig.~\ref{chi_cont_var}.
The P-wave tensor splitting (Fig.~\ref{1Pte}) receives heavy-quark discretization effects from the same mass mismatches as the 1S
hyperfine splitting, and the observed total discretization effects in the 1P tensor splitting are of the same absolute size as those
in the 1S hyperfine splitting.
Variations of our fit model are displayed in the fourth pane of Fig.~\ref{chi_cont_var}.
As in the case of the hyperfine splitting, we take the difference between variation B (omitting the coarsest
ensembles) and the default fit model as an additional systematic uncertainty.

\begin{figure}[tbp]
\centering
\includegraphics[clip,width=10.0cm]{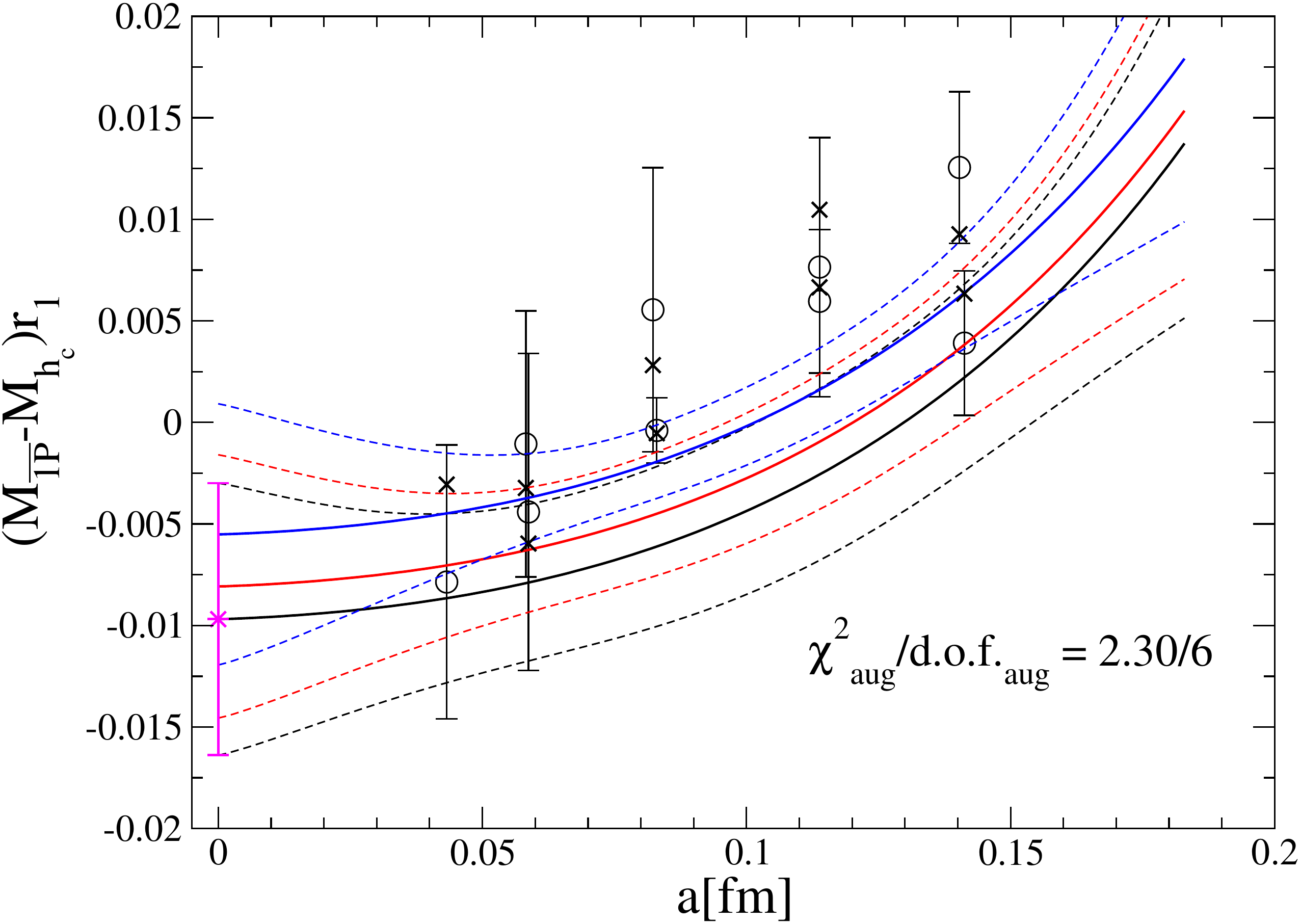}
\caption{Chiral and continuum fit for the P-wave hyperfine splitting.
    The black circles denote the lattice data. Curves for physical (black), $0.1m_s$, (red) and $0.2m_s$ (blue) light-quark masses 
    are plotted.
    The black crosses show the fit results evaluated at the lattice parameters of the gauge ensemble.
    The magenta symbol indicates the result in the combined chiral and continuum limit.}
\label{1Phf}
\end{figure}

The 1P hyperfine splitting defined in Eq.~(\ref{eq:1PHF}) is expected to be very small and, indeed, experiments measure a value
compatible with zero.
Our results are shown in Fig.~\ref{1Phf}.
Our data for this quantity are rather noisy.
We find a central value slightly more than $1\sigma$ away from zero, but we do not believe this extrapolation to be fully under
control, and the strong cancellation may make this combination sensitive to charm-anticharm annihilation.

\subsection{2S-1S splitting}

Beyond the mesons listed in Table~\ref{lowstates}, the only known charmonia below the $\bar{D}D$ threshold are the $\psi(2S)$ and
$\eta_c(2S)$.
With our interpolator basis these states are not well determined.
Furthermore, we do not include the $\bar{D}D$ scattering states in our basis
and the threshold states therefore cannot be cleanly
separated from the close-to-threshold 2S bound states.
As a result, the energy values we obtain depend strongly on the lower boundary $t_\text{min}$ of the fit range, as demonstrated
previously in~\cite{DeTar:2012xk}.
This issue is not seen in a recent simulation of the $\psi(3770)$ resonance using a more sophisticated basis of both quark-antiquark
and $\bar{D}D$ interpolators~\cite{Lang:2015sba}, where the QCD bound state corresponding to the $\psi(2S)$ can be obtained to a
good statistical precision.
Note, however, that Ref.~\cite{Lang:2015sba} was limited to just two sets of
gauge configurations, so that the chiral and continuum limits could not be taken.

\subsection{Uncertainty estimates}
\label{uncertainty}

\begin{figure}[tbp]
\centering
\includegraphics[clip,width=10.0cm]{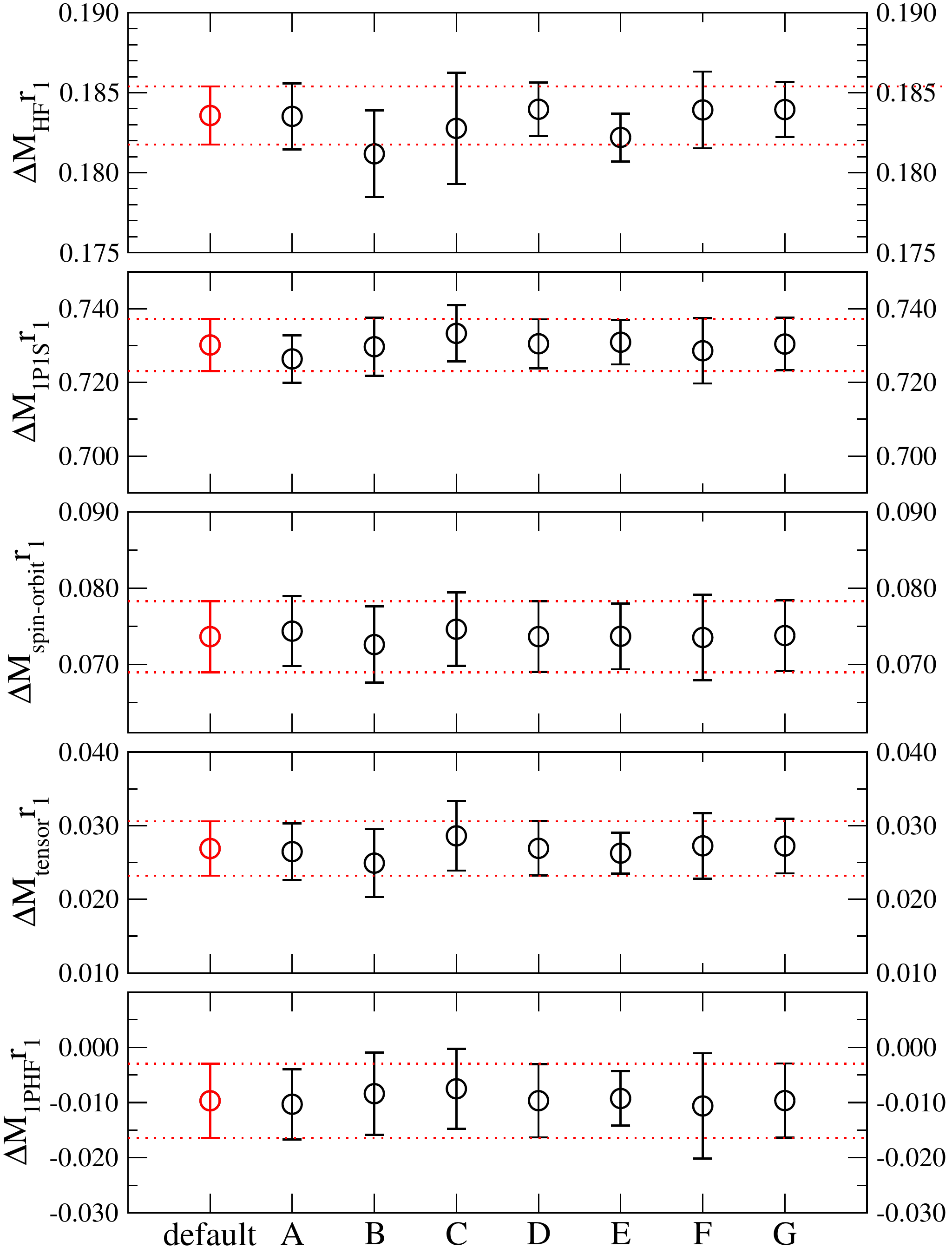}
\caption{Systematic variation of the charmonium mass splittings when varying
  the details of the chiral-continuum extrapolation. The label ``default'' indicates our
  final result described in detail for each splitting in Sec.~\ref{results}.
  The variations (letters A to G) are described in Table~\ref{fitvariations}.}
\label{chi_cont_var}
\end{figure}

To obtain final best estimates for the calculated mass splittings, we need to
assess the relevant systematic uncertainties associated with our procedures.
In total, we consider uncertainties arising from correlator fits, from the
charm-quark mass tuning procedure, from the correction to physical charm-quark
mass described in Sec.~\ref{sec:correction}, from the chiral-continuum fit,
and from our limited knowledge of the lattice scale. For the charm-quark
tuning procedure, the main uncertainty in the continuum limit arises from the
effect of the lattice-scale determination on the charm-quark tuning.
We account for this uncertainty as part of our scale-setting uncertainty below.
All relevant uncertainties are tabulated in Table~\ref{errorbudget}.
We now discuss them in turn.

\subsubsection{Variations of the correlator fits}

We have investigated many variations of the correlator fits, including
correlator basis variations, variations of the fit interval, fit shape, etc.,
and found that our results are stable under sensible variations of the fitting procedure.
Our final choices, which are displayed in full in Tables~\ref{swave_table} and~\ref{pwave_table} (in
Appendix~\ref{app:levels}), are reasonably conservative and encompass almost
all stable fit choices with a reasonable goodness of fit. For our final
uncertainty estimate, we therefore do not include an additional uncertainty
for these variations in correlator fits.

\subsubsection{Uncertainty from the determination of the slope in $\kappa_c$}

Variation ``G'' from Table~\ref{fitvariations} illustrates the results when the slope in $\kappa_c$ used for the charm-quark mass
corrections is varied by one standard deviation.
This variation is small and straightforward to quantify.

\subsubsection{Variations of the chiral-continuum fits}

Beyond our default fit results, Table~\ref{fitvariations} lists several variations of the chiral-continuum fit we performed.
The results associated with these variations are shown together with our default results in Fig.~\ref{chi_cont_var}.
Among these, variations A--D vary the fit forms used, while variations E and~F
test whether the results are sensitive
to the prior widths selected for the coefficients of the
heavy-quark-discretization shapes from Fig.~\ref{shapes}.
In general the variations among the different fits are rather mild.
Significant variations have been discussed for each observables in the previous subsections.

For the S-wave hyperfine splitting and the P-wave tensor splitting, we assess
the systematic uncertainty of the chiral-continuum fit by taking the
difference between the default fit and variation ``B'', which results from
omitting our data at the coarsest lattice spacing. For all other splittings,
the variations are insignificant compared with the statistical uncertainty of the
fit. While wider priors leads to a slightly increased uncertainty
estimate, there is no significant variation in our best estimates for the
splittings.

\subsubsection{Uncertainty from the determination of the lattice scale}

The procedure for our determination of the scale-setting uncertainty is described above in Sec.~\ref{scale_error}.
For the 1S-hyperfine and 1P-1S splittings this uncertainty is of the same size as the statistical uncertainty from
the chiral-continuum fit.
In particular, the indirect uncertainty stemming from the uncertainty of the determination of the charm-quark hopping parameter
$\kappa_c$ on the scale setting is quite large and this uncertainty has been neglected in some of the literature.
For the spin-orbit splitting the uncertainty from scale setting is somewhat
smaller than the statistical uncertainty after extrapolation to the physical point.
The scale-setting uncertainties for the other splittings are small.

\begin{table}[tbp]
\centering
\caption{Systematic uncertainties on the mass splittings in MeV. An asterisk (*)
  indicates that the corresponding uncertainty is small compared with the
  statistical uncertainty of the chiral-continuum fit and can therefore be
  neglected in quantifying the total uncertainty. For the total systematic
  uncertainty we add the single values in quadrature.
  Recall that our simulation omits charm-anticharm annihilation.}
\label{errorbudget}
\begin{tabular}{lccccc}
\hline
\hline
 \T\B Source & 1P-1S & 1S hyperfine & 1P spin-orbit & 1P
tensor & 1P hyperfine \\
\hline
Slope  in $\kappa_c$ & 0.2 & 0.2 & 0.1 & 0.2 & (*)\\
Chiral-continuum fit shape & (*) & 1.5 & (*) & 1.6 & (*)\\
Lattice scale        & 3.3 & 1.6 & 0.9 & 0.1 & (*)\\
\hline
Total & 3.3 & 2.2 & 0.9 & 1.6 & $<$0.1\\
\hline
\hline
\end{tabular}
\end{table}

\subsection{Comparison with previous calculations}
\label{comp_prev}

The Fermilab Lattice and MILC collaborations have previously reported results for the mass splittings in the low-lying charmonium
spectrum~\cite{Burch:2009az}.
Our current results use the same library of gauge configurations.
Compared with the previous study we make use of finer lattice spacings, a better determination of the physical quark masses (in
particular an improved determination of the charm-quark hopping parameter $\kappa_c$~\cite{Bailey:2014tva}) and of the lattice
spacings used in the simulation~\cite{Bazavov:2011aa}.
All these ingredients allowed us to perform a more sophisticated chiral-continuum extrapolation.
The new results supersede those of Ref.~\cite{Burch:2009az}.

\begin{table}[tbp]
    \centering
    \caption{Charmonium mass splittings obtained in this paper compared with the calculation on the same library of gauge-field
        configurations from~\cite{Burch:2009az}.
        For an explanation of differences between the two calculations please refer to the text.
        The quoted uncertainties are statistical and systematic; the third uncertainty on the 1S hyperfine splitting is the estimate
        for the downward shift due to disconnected contributions from Ref.~\cite{Levkova:2010ft}.}
    \label{compareFNAL}
\begin{tabular}{lcc}
\hline
\hline
 \T\B Mass difference & This analysis [MeV] & Ref.~\cite{Burch:2009az} [MeV]\\
\hline
1S hyperfine & $116.2\pm1.1 \pm2.2 {}^{-1.5}_{-4.0}$ & $116.0\pm 7.4^{+2.6}_{-0}$\\
1P-1S splitting & $462.2\pm4.5 \pm3.3$ & $473\pm 12^{+10}_{-0}$\\
1P spin-orbit & $46.6\pm3.0 \pm0.9$ & $43.3\pm 6.6^{+1.0}_{-0}$ \\
1P tensor & $17.0\pm2.3 \pm1.6$ &  $15.0\pm 2.3^{+0.3}_{-0}$ \\
1P hyperfine & $-6.1\pm4.2 \pm 0.1$ & -- \\
\hline
\hline
\end{tabular}
\end{table}

Table~\ref{compareFNAL} shows a direct comparison of the previous results from~\cite{Burch:2009az} with our new results.
With the exception of the 1P tensor splitting all the new results are quite a bit more precise.
For the 1P tensor splitting our more elaborate chiral-continuum extrapolation leads to a significant increase in the estimate of the
associated uncertainty; the previously quoted uncertainty was probably underestimated.

\begin{figure}[tbp]
    \centering
    \includegraphics[clip,width=10.0cm]{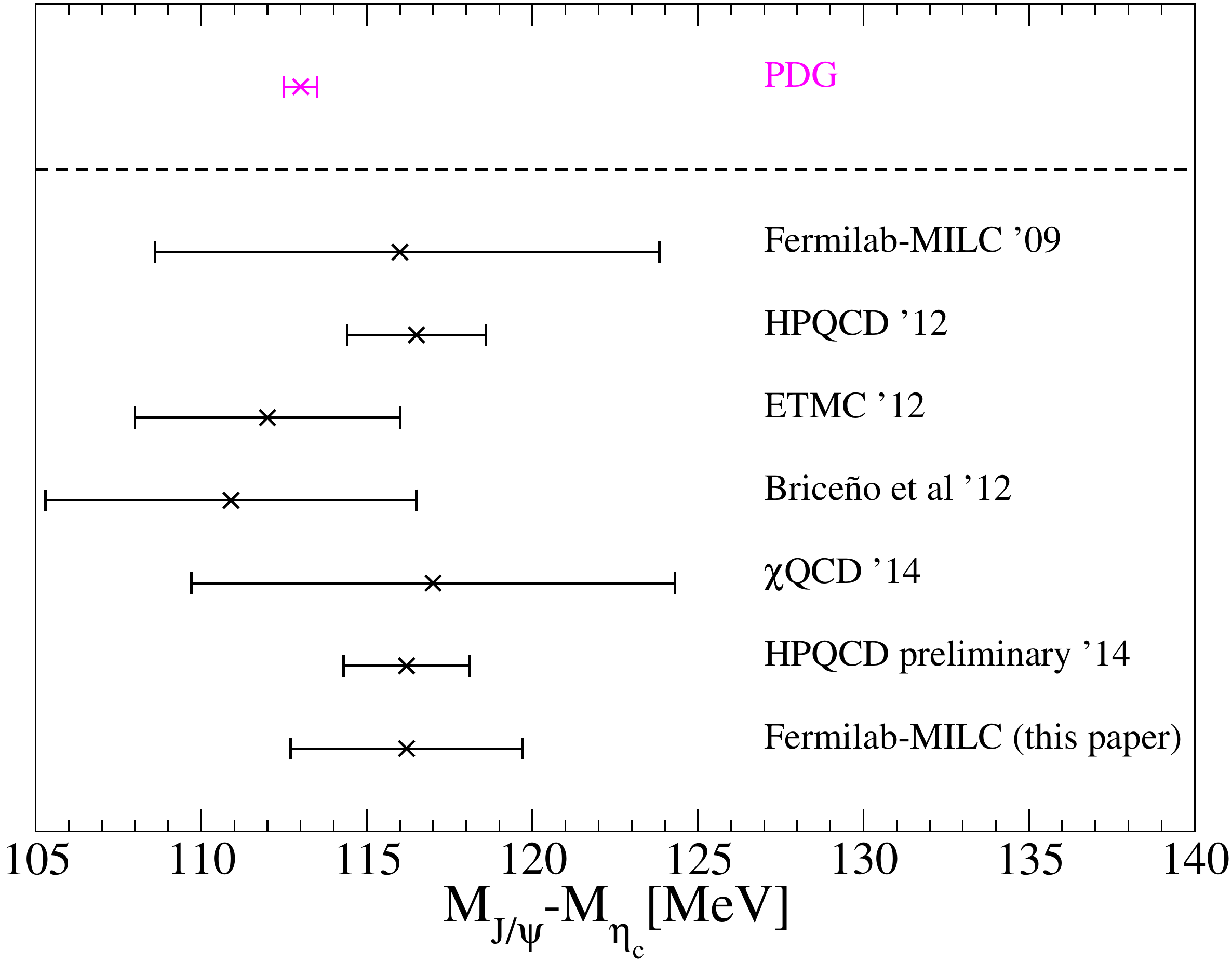}
    \caption{The connected part of the 1S hyperfine splitting from lattice-QCD calculations that include a continuum limit (and an
        extrapolation to physical sea-quark masses where appropriate).
        Effects from charm-anticharm annihilation (of valence quarks) are not included in any of these calculations.
        For comparison we also show the 1S hyperfine splitting from the PDG~\cite{Tanabashi:2018oca}.
        Note that the lattice calculations neglecting disconnected contributions and treating the $\eta_c$ as stable need not
        result in the same value as the PDG.}
    \label{hyperfine_comparison}
\end{figure}

For the 1S hyperfine splitting there have been several lattice simulations aimed at a full control of systematic uncertainties in
the QCD calculation of the connected contribution to the hyperfine
splitting~\cite{Burch:2009az,Donald:2012ga,Becirevic:2012dc,Briceno:2012wt,Yang:2014sea}.
Preliminary results with simulations using the HISQ action for charm quarks have also been presented in
Ref.~\cite{Galloway:2014tta}.
In particular, all these references quote results for physical sea-quark masses in the continuum limit.
Figure~\ref{hyperfine_comparison} shows a visual comparison of these calculations.
The results from various collaborations are quite consistent.
Unfortunately, all these results neglect effects from annihilation of the valence charm-quarks.
These have previously been estimated from lattice QCD~\cite{Levkova:2010ft} and from perturbation theory~\cite{Donald:2012ga}.
Note that these results disagree in the sign of the annihilation effects.

More importantly, charm-anticharm annihilation in the physical system results in a substantial total
hadronic width of the $\eta_c$, dominating the total width of $32.0(8)$~MeV~\cite{Tanabashi:2018oca}.
The PDG lists many decays both into hadronic resonances and into stable final states~\cite{Tanabashi:2018oca}.
While lattice QCD studies of hadronic resonances using L\"uscher's finite volume method~\cite{Luscher:1990ux,Luscher:1991cf} are
continuing to make considerable progress (for a review see Ref.~\cite{Briceno:2017max}) and are now being applied to states close to
double open charm
thresholds~\cite{Prelovsek:2013cra,Lee:2014uta,Lang:2015sba,Cheung:2017tnt}, a
rigorous study of the $\eta_c$ on the lattice is currently out of reach.%
\footnote{For interesting new developments concerning the extraction of total
  decay rates into multi hadron final states please
refer to Ref.~\cite{Hansen:2017mnd}.} %
In our current calculation, the uncertainty from neglecting disconnected contributions is now the largest
uncertainty in the error budget for the 1S hyperfine splitting.

\section{Conclusions and Outlook}
\label{conclusions}

\begin{table}[tbp]
\centering
\caption{
  Charmonium mass splittings compared with the experimental
  values. The quoted uncertainties are statistical and systematic, where the
  systematic uncertainty is discussed in Sec.~\ref{uncertainty}. Note that
  our simulation neglects charm-quark annihilation diagrams. The
  second systematic uncertainty on the 1S hyperfine splitting is best-estimate
  for the downward shift due to such disconnected contributions~\cite{Levkova:2010ft}.}
\label{numbers}
\begin{tabular}{lcc}
\hline
\hline
 \T\B Mass difference & This analysis [MeV] & Experiment [MeV]\\
\hline
1S hyperfine & $116.2 \pm 1.1 \pm3.3 {}^{-1.5}_{-4.0}$ & $113.0\pm0.5$\\
1P-1S splitting & $462.2\pm 4.5 \pm3.3$ & $456.64\pm0.14$\\
1P spin-orbit & $46.6\pm 3.0 \pm0.9 $ &  $46.60\pm0.08$ \\
1P tensor & $17.0\pm 2.3 \pm1.6 $ &  $16.27\pm0.07$ \\
1P hyperfine & $-6.1 \pm 4.2 \pm 0.1 $ & $-0.09\pm0.14$ \\
\hline
\hline
\end{tabular}
\end{table}

In this paper, we have presented results for the splittings of low-lying charmonium states.
Table~\ref{numbers} shows a comparison of our results with the experimental values~\cite{Tanabashi:2018oca}.
Within our uncertainty estimates, which are described in detail in Sec.~\ref{uncertainty}, the lattice QCD postdictions
are in excellent agreement with experiment, demonstrating that heavy-quark discretization effects for charmonium are well controlled
in our setup.
Our results improve upon previous results by the Fermilab Lattice and MILC collaborations presented in Ref.~\cite{Burch:2009az}, which are now superseded.

While our determination of the 1S hyperfine splitting uses the estimate for the charm-annihilation contribution from
Ref.~\cite{Levkova:2010ft}, all current lattice determination including the results presented here neglect effects from
charm-anticharm annihilation.
For the 1S hyperfine splitting this is now the largest source of uncertainty.
A possible direction of further research in this context would be a precision study and prediction of spin-splittings in the $B_c$
system, where the contributions from annihilation diagrams are absent.
Note that the the hyperfine splitting between the $B_c^*$ and $B_c$ mesons has already been predicted from lattice QCD in
\cite{Gregory:2009hq,Dowdall:2012ab,Mathur:2018epb}, while the $B_c^*$ has not yet been seen in experiment.

\acknowledgments

Computation for this work was done at the Argonne Leadership Computing Facility (ALCF), Blue~Waters at the National Center for
Supercomputing Applications (NCSA), the National Energy Resources Supercomputing Center (NERSC), the National Institute for
Computational Sciences (NICS), the Texas Advanced Computing Center (TACC), and the USQCD facilities at Fermilab, under grants from
the NSF and DOE.
C.D. and S.-H.L.  are supported by the U.S.\ National Science Foundation under grants Nos.~PHY09-03571 and
PHY14-14614, and the U.S.\ Department of Energy under grant No.~DE-FC02-12ER41879.
A.S.K.\ acknowledges support by the German Excellence Initiative and the European Union Seventh Framework Programme under grant
agreement No.~291763 as well as the European Union's Marie Curie COFUND program.
This document was prepared by the Fermilab Lattice and MILC Collaborations using the resources of the Fermi National Accelerator
Laboratory (Fermilab), a U.S.\ Department of Energy, Office of Science, HEP User Facility.
Fermilab is managed by Fermi Research Alliance, LLC (FRA), acting under Contract No.\ DE-AC02-07CH11359.


\appendix

\section{Tables of interpolators}
\label{interpolators}

Table~\ref{interpolator_table} provides the interpolators in each irreducible representation of the (lattice) cubic group, parity
$P$, and charge conjugation $C$ quantum numbers.
Entries correspond to the $O_i$ in Eqs.~(\ref{eq:bilinears}), and the smearing types $\nabla_i$, $\mathbb{D}_i$, and $\mathbb{B}_i$
are defined in Eqs.~(\ref{operators}).

\begin{table*}[tbp]
    \centering
    \caption{Schematic of interpolators for each lattice irreducible representation.
    Repeated indices are summed over.
    Interpolators without derivatives are used with both stochastic Gaussian-smeared (G) and stochastic point (P) sources
    and sinks, as detailed in Sec.~\ref{observables}.
    $\gamma_t$ denotes the $\gamma$ matrix in time direction, and the Clebsch-Gordan coefficients $Q_{ijk}$ are given in reference
    Ref.~\cite{Dudek:2007wv}.}
    \label{interpolator_table}
\begin{tabular}{ccccccc}
\hline
\hline
$A_1^{-+}$ & $A_1^{++}$ & $T_1^{--}$ &$T_1^{+-}$ & $T_1^{++}$ & $T_2^{++}$ & $E^{++}$ \\ \hline
$\gamma_5$ (G) & $\mathds{1}$ (G) & $\gamma_i$ (G) & $\gamma_t\gamma_5\gamma_i$ (G) & $\gamma_5\gamma_i$ (G) & $|\varepsilon_{ijk}|\gamma_j\nabla_k$ & $Q_{ijk}\gamma_j\nabla_k$\\
$\gamma_5$ (P) & $\mathds{1}$ (P) & $\gamma_i$ (P) & $\gamma_t\gamma_5\gamma_i$ (P) & $\gamma_5\gamma_i$ (P) & $|\varepsilon_{ijk}|\gamma_t\gamma_j\nabla_k$ &  $Q_{ijk}\gamma_t\gamma_j\nabla_k$\\
$\gamma_t\gamma_5$ (G) &  $\gamma_i\nabla_i$ & $\gamma_t\gamma_i$ (G) & $\gamma_5\nabla_i$ & $\varepsilon_{ijk}\gamma_j\nabla_k$ & $\mathbb{D}_i$ &  $Q_{ijk}\gamma_5\gamma_j\mathbb{D}_k$\\
$\gamma_t\gamma_5$ (P) & $\gamma_t\gamma_i\nabla_i$ & $\gamma_t\gamma_i$ (P) & $\gamma_t\gamma_5\nabla_i$ & $\varepsilon_{ijk}\gamma_t\gamma_j\nabla_k$ & $|\varepsilon_{ijk}|\gamma_t\gamma_5\gamma_j\mathbb{B}_k$ & $Q_{ijk}\gamma_t\gamma_5\gamma_j\mathbb{B}_k$\\
$\gamma_t\gamma_5\gamma_i\nabla_i$ & $\gamma_t\gamma_5\gamma_i\mathbb{B}_i$ & $\nabla_i$ & $|\varepsilon_{ijk}|\gamma_t\gamma_5\gamma_j\mathbb{D}_k$ & $|\varepsilon_{ijk}|\gamma_5\gamma_j\mathbb{D}_k$ & & \\
$\gamma_i\mathbb{B}_i$ & & $\varepsilon_{ijk}\gamma_5\gamma_j\nabla_k$ &  $\mathbb{B}_i$ & $\gamma_t\mathbb{B}_i$ & &\\
$\gamma_t\gamma_i\mathbb{B}_i$ & & $|\varepsilon_{ijk}|\gamma_j\mathbb{D}_k$ & $\varepsilon_{ijk}\gamma_5\gamma_j\mathbb{B}_k$ & $\varepsilon_{ijk}\gamma_t\gamma_5\gamma_j\mathbb{B}_k$ &&\\
& &  $|\varepsilon_{ijk}|\gamma_t\gamma_j\mathbb{D}_k$ &&&&\\
& &  $\gamma_5\mathbb{B}_i$ &&&&\\
& & $\gamma_t\gamma_5\mathbb{B}_i$ &&&&\\
\hline
\hline
\end{tabular}
\end{table*}

\clearpage

\section{Ground-state energy levels}
\label{app:levels}

Tables~\ref{swave_table} and~\ref{pwave_table} list the determined ground
state masses for each ensemble and quantum number combination along with the interpolator
basis, reference timeslice $t_0$ of the variational method, fit range and fit
type used to obtain the result.

\begin{table*}[th]
\centering
\caption{Mass $aM$ for the 1S states on all ensembles.
    The ensembles are labeled by their lattice spacing $a$ and ratio of sea quark masses $m_l^\prime/m_s^\prime$.
    The basis of interpolators is labeled according to Table~\ref{interpolator_table}.
    All fits are two-exponential fits in the specified fit range.
    As we only analyze the autocorrelation within the Monte-Carlo chain for the mass splittings the printed $\chi^2$ per degree of
    freedom is somewhat larger than one.
    For our final results autocorrelations have been taken into account where necessary.}
\label{swave_table}
\begin{tabular}{ccS[table-format=1.5]ccccS[table-format=1.5]c}
\hline
\hline
 \T\B $\approx$a [fm]& $m_l^\prime/m_s^\prime$ & $\kappa_\text{sim}$ & $J^{PC}$ & $t_0$ & basis & fit range & $aM$ &
 $\chi^2/\mathrm{d.o.f.}$\\
\hline
0.14 & 0.2 & 0.1221  & $0^{-+}$ &2& 1,2,3,4,5 & 2--20& 1.67622( 15)&0.84\\
0.14 & 0.1 & 0.1221  & $0^{-+}$ &2& 1,2,3,4,5 & 2--20& 1.67746( 10)&2.35\\
0.114& 0.2 & 0.12423 & $0^{-+}$ &2& 1,2,3,4,5 & 3--27& 1.46897(  8)&1.50\\
0.114& 0.1 & 0.1220  & $0^{-+}$ &2& 1,2,3,4,5 & 3--27& 1.58055(  7)&0.88\\
0.114& 0.1 & 0.1245  & $0^{-+}$ &2& 1,2,3,4,5 & 3--27& 1.45278(  8)&1.06\\
0.114& 0.1 & 0.1280  & $0^{-+}$ &2& 1,2,3,4,5 & 3--27& 1.26162(  9)&1.70\\
0.082& 0.2 & 0.12722 & $0^{-+}$ &3& 1,2,3,4,5 & 4--42& 1.14427(  8)&1.20\\
0.082& 0.1 & 0.12714 & $0^{-+}$ &3& 1,2,3,4,5 & 4--42& 1.15211(  4)&1.31\\
0.058& 0.2 & 0.1298  & $0^{-+}$ &5& 1,2,3,4,5 & 6--64& 0.83119(  4)&1.32\\
0.058& 0.1 & 0.1296  & $0^{-+}$ &5& 1,2,3,4,5 & 6--64& 0.84756(  2)&1.08\\
0.043& 0.2 & 0.1310  & $0^{-+}$ &6& 1,2,3,4,5 & 8--81& 0.63519(  3)&1.60\\
\hline
0.14 & 0.2 & 0.1221  & $1^{--}$ &2& 1,2,5,6,7,8 & 2--20& 1.75238( 22)&0.51\\
0.14 & 0.1 & 0.1221  & $1^{--}$ &2& 1,2,5,6,7,8 & 2--20& 1.75324( 16)&1.64\\
0.114& 0.2 & 0.12423 & $1^{--}$ &2& 1,2,5,6,7,8 & 3--27& 1.53353( 14)&1.46\\
0.114& 0.1 & 0.1220  & $1^{--}$ &2& 1,2,5,6,7,8 & 3--27& 1.63834( 13)&0.92\\
0.114& 0.1 & 0.1245  & $1^{--}$ &2& 1,2,5,6,7,8 & 3--27& 1.51690( 13)&1.06\\
0.114& 0.1 & 0.1280  & $1^{--}$ &2& 1,2,5,6,7,8 & 3--27& 1.33715( 18)&1.31\\
0.082& 0.2 & 0.12722 & $1^{--}$ &3& 1,2,5,6,7,8 & 4--42& 1.19131( 20)&1.53\\
0.082& 0.1 & 0.12714 & $1^{--}$ &3& 1,2,5,6,7,8 & 4--42& 1.19873(  7)&1.01\\
0.058& 0.2 & 0.1298  & $1^{--}$ &5& 1,2,5,6,7,8 & 6--64& 0.86508(  9)&1.32\\
0.058& 0.1 & 0.1296  & $1^{--}$ &5& 1,2,5,6,7,8 & 6--64& 0.88092(  5)&1.18\\
0.043& 0.2 & 0.1310  & $1^{--}$ &6& 1,2,5,6,7,8 & 8--81& 0.66053(  6)&1.66\\
\hline
\hline
\end{tabular}
\end{table*}

\clearpage

\setlength{\LTcapwidth}{\textwidth}
\begin{longtable}{ccS[table-format=1.5]ccccS[table-format=1.5]c}
\caption{Mass $aM$ for the 1P states on all ensembles. The ensembles are
  labeled by their lattice spacing $a$ and ratio of sea quark masses
  $m_l^\prime/m_s^\prime$. For $J^{PC}=2^{++}$ results from two lattice irreducible
  representations ($T_2$ and $E$) are listed. For further comments see Table~\ref{swave_table}.}
\label{pwave_table} \\
\hline
\hline
\T\B $\approx$a [fm]& $m_l^\prime/m_s^\prime$ & $\kappa_\text{sim}$ & $J^{PC}$ & $t_0$ & basis & fit range & $aM$ &
$\chi^2/\mathrm{d.o.f.}$\\
\hline
0.14  & 0.2 & 0.1221 & $0^{++}$ & 3& 1,2,3,4 &3--10& 2.0436( 31)&0.22\\
0.14  & 0.1 & 0.1221 & $0^{++}$ &3& 1,2,3,4 &3--10& 2.0368( 35)&0.33\\
0.114 & 0.2 & 0.12423 & $0^{++}$ &3& 1,2,3,4 &3--16& 1.7556( 33)&0.65\\
0.114 & 0.1 & 0.1220 & $0^{++}$ &3& 1,2,3,4 &3--12& 1.8651( 13)&0.11\\
0.114 & 0.1 & 0.1245 & $0^{++}$ &3& 1,2,3,4 &3--12& 1.7390( 15)&0.72\\
0.114 & 0.1 & 0.1280 & $0^{++}$ &3& 1,2,3,4 &3--12& 1.5220(928)&0.87\\
0.082 & 0.2 & 0.12722 & $0^{++}$ &3& 1,2,3,4 &3--19& 1.3421( 13)&0.23\\
0.082 & 0.1 & 0.12714 & $0^{++}$ &3& 1,2,3,4 &3--19& 1.3481(  7)&0.45\\
0.058 & 0.2 & 0.1298 & $0^{++}$ &5& 1,2,3,4 &6--28& 0.9646( 10)&0.77\\
0.058 & 0.1 & 0.1296 & $0^{++}$ &5& 1,2,3,4 &6--30& 0.9807( 10)&0.57\\
0.043 & 0.2 & 0.1310 & $0^{++}$ &6& 1,2,3,4 &7--31& 0.7349(  8)&0.98\\
\hline
0.14  & 0.2 & 0.1221 & $1^{++}$ &3& 1,2,3,4,5 &3--10& 2.0896( 28)&0.20\\
0.14  & 0.1 & 0.1221 & $1^{++}$ &3& 1,2,3,4,5 &3--10& 2.0854( 22)&0.55\\
0.114 & 0.2 & 0.12423 & $1^{++}$ &3& 1,2,3,4,5 &3--16& 1.8017( 31)&0.69\\
0.114 & 0.1 & 0.1220 & $1^{++}$ &3& 1,2,3,4,5 &3--12& 1.9039( 18)&0.61\\
0.114 & 0.1 & 0.1245 & $1^{++}$ &3& 1,2,3,4,5 &3--17& 1.7800( 29)&0.45\\
0.114 & 0.1 & 0.1280 & $1^{++}$ &3& 1,2,3,4,5 &6--14& 1.5898(338)&0.20\\
0.082 & 0.2 & 0.12722 & $1^{++}$ &3& 1,2,3,4,5 &3--23& 1.3783( 24)&1.29\\
0.082 & 0.1 & 0.12714 & $1^{++}$ &3& 1,2,3,4,5 &3--24& 1.3843(  8)&1.14\\
0.058 & 0.2 & 0.1298 & $1^{++}$ &5& 1,2,3,4,5 &6--24& 0.9930( 12)&0.93\\
0.058 & 0.1 & 0.1296 & $1^{++}$ &5& 1,2,3,4,5 &6--32& 1.0070( 11)&0.81\\
0.043 & 0.2 & 0.1310 & $1^{++}$ &6& 1,2,3,4,5 &7--41& 0.7542( 10)&0.68\\
\hline
0.14  & 0.2 & 0.1221 & $2^{++} (T_2)$ &3& 1,2,3,4 &3--10& 2.1243( 39)&0.31\\
0.14  & 0.1 & 0.1221 & $2^{++} (T_2)$ &3& 1,2,3,4 &3--10& 2.1193( 36)&1.34\\
0.114 & 0.2 & 0.12423 & $2^{++} (T_2)$ &3& 1,2,3,4 &3--16& 1.8304( 24)&0.64\\
0.114 & 0.1 & 0.1220 & $2^{++} (T_2)$ &3& 1,2,3,4 &3--19& 1.9291( 23)&0.48\\
0.114 & 0.1 & 0.1245 & $2^{++} (T_2)$ &3& 1,2,3,4 &3--18& 1.8100( 23)&0.24\\
0.114 & 0.1 & 0.1280 & $2^{++} (T_2)$ &3& 1,2,3,4 &3--17& 1.5045(707)&0.64\\
0.082 & 0.2 & 0.12722 & $2^{++} (T_2)$ &3& 1,2,3,4 &3--19& 1.4006( 25)&0.79\\
0.082 & 0.1 & 0.12714 & $2^{++} (T_2)$ &3& 1,2,3,4 &3--23& 1.4045( 11)&0.79\\
0.058 & 0.2 & 0.1298 & $2^{++} (T_2)$ &5& 1,2,3,4 &6--25& 1.0082( 19)&0.92\\
0.058 & 0.1 & 0.1296 & $2^{++} (T_2)$ &5& 1,2,3,4 &6--29& 1.0176( 42)&1.05\\
0.043 & 0.2 & 0.1310 & $2^{++} (T_2)$ &6& 1,2,3,4 &7--31& 0.7644( 20)&0.81\\
\hline
0.14  & 0.2 & 0.1221 & $2^{++} (E)$ &3& 1,2,3,4 &3--10& 2.1259( 37)&0.12\\
0.14  & 0.1 & 0.1221 & $2^{++} (E)$ &3& 1,2,3,4 &3--10& 2.1190( 39)&1.50\\
0.114 & 0.2 & 0.12423 & $2^{++} (E)$ &3& 1,2,3,4 &3--13& 1.8289( 38)&0.23\\
0.114 & 0.1 & 0.1220 & $2^{++} (E)$ &3& 1,2,3,4 &3--17& 1.9289( 24)&0.70\\
0.114 & 0.1 & 0.1245 & $2^{++} (E)$ &3& 1,2,3,4 &3--15& 1.8115( 20)&0.23\\
0.114 & 0.1 & 0.1280 & $2^{++} (E)$ &3& 1,2,3,4 &3--17& 1.5935(437)&0.67\\
0.082 & 0.2 & 0.12722 & $2^{++} (E)$ &3& 1,2,3,4 &3--19& 1.4018( 20)&0.98\\
0.082 & 0.1 & 0.12714 & $2^{++} (E)$ &3& 1,2,3,4 &3--23& 1.4051( 11)&0.87\\
0.058 & 0.2 & 0.1298 & $2^{++} (E)$ &5& 1,2,3,4 &6--29& 1.0081( 21)&0.89\\
0.058 & 0.1 & 0.1296 & $2^{++} (E)$ &5& 1,2,3,4 &6--24& 1.0201( 35)&1.57\\
0.043 & 0.2 & 0.1310 & $2^{++} (E)$ &6& 1,2,3,4 &7--31& 0.7633( 26)&0.84\\
\hline
0.14  & 0.2 & 0.1221 & $1^{+-}$ &3& 1,2,3,4,5 &3--10& 2.0986( 33)&0.08\\
0.14  & 0.1 & 0.1221 & $1^{+-}$ &3& 1,2,3,4,5 &3--10& 2.0970( 27)&1.25\\
0.114 & 0.2 & 0.12423 & $1^{+-}$ &3& 1,2,3,4,5 &3--16& 1.8093( 35)&0.22\\
0.114 & 0.1 & 0.1220 & $1^{+-}$ &3& 1,2,3,4,5 &3--16& 1.9116( 23)&0.52\\
0.114 & 0.1 & 0.1245 & $1^{+-}$ &3& 1,2,3,4,5 &3--16& 1.7903( 23)&0.58\\
0.114 & 0.1 & 0.1280 & $1^{+-}$ &3& 1,2,3,4,5 &3--17& 1.5965(102)&0.43\\
0.082 & 0.2 & 0.12722 & $1^{+-}$ &3& 1,2,3,4,5 &3--21& 1.3856( 29)&1.96\\
0.082 & 0.1 & 0.12714 & $1^{+-}$ &3& 1,2,3,4,5 &3--21& 1.3917(  8)&0.78\\
0.058 & 0.2 & 0.1298 & $1^{+-}$ &5& 1,2,3,4,5 &6--35& 0.9985( 19)&0.87\\
0.058 & 0.1 & 0.1296 & $1^{+-}$ &5& 1,2,3,4,5 &6--35& 1.0115( 13)&0.84\\
0.043 & 0.2 & 0.1310 & $1^{+-}$ &6& 1,2,3,4,5 &7--42& 0.7585( 11)&1.21\\
\hline
\hline
\end{longtable}

\bibliography{asqtad_onium}
\bibliographystyle{apsrev4-1}

\end{document}